# Stacking faults from a different angle – Overcoming the edge-on limit in high-resolution defect analysis

**Nicolas Karpstein [a], Lukas Müller [a], Andreas Bezold [b,c], Michael J. Mills [c], Steffen Neumeier [b], Erdmann Spiecker [a,*]**

[a] Friedrich-Alexander-Universität Erlangen-Nürnberg, Department of Materials Science & Engineering, Institute of Micro- and Nanostructure Research, and Center for Nanoanalysis and Electron Microscopy (CENEM), IZNF, Erlangen, Germany

[b] Friedrich-Alexander-Universität Erlangen-Nürnberg, Department of Materials Science & Engineering, Institute I: General Materials Properties, Erlangen, Germany

[c] The Ohio State University, Department of Materials Science & Engineering, Columbus, OH, USA

* Corresponding author: erdmann.spiecker@fau.de

---

## Abstract

The nature of stacking faults (SF)—whether intrinsic or extrinsic—plays a pivotal role in defect-mediated processes in crystalline materials. Yet, transmission electron microscopy (TEM) techniques for their reliable analysis remain limited to either conventional fringe-contrast imaging of inclined faults or atomic-resolution imaging of edge-on configurations. Here, we overcome this long-standing geometric constraint by introducing a high-resolution scanning TEM method that enables full structural discrimination of inclined SFs, as demonstrated in fcc, $L1_2$, and sphalerite crystals. This approach complements edge-on analysis and provides access to SFs on all glide planes along commonly utilized zone axes. We demonstrate the method's robustness in a CoNi-based superalloy by discriminating fault types even in overlapping configurations and for foil thicknesses exceeding 100 nm, and extend its application to analyzing bounding partial dislocations in inclined geometries to reveal the fault formation mechanism. Probe propagation simulations reveal that fault-induced de-channeling is key to contrast formation and is strongly governed by the fault's depth within the sample. Leveraging this effect, we further establish a route to artificially generate ultrathin TEM

lamellae—bounded by the SF itself—enhancing contrast for atomic-scale studies of long-range ordering and compositional fluctuations.

---

## Introduction

Stacking faults (SFs) are among the most significant defects in crystalline materials. Since their initial theoretical description and subsequent experimental observation by X-ray diffraction around 1940 [1, 2], they have been identified in a broad range of materials, including metals [3], layered crystals [4], and semiconductors [5, 6]. Far from being mere imperfections, SFs can profoundly influence both the structural and functional properties of materials (e.g. [5, 7, 8]). While SFs often originate from dislocation glide, they can also arise from the agglomeration of point defects associated with dislocation climb, or form directly during the crystal growth process. Given their diverse origins and significant impact on material behavior, it is therefore essential to characterize SFs reliably and comprehensively across a range of experimental conditions.

SFs are particularly prevalent in close-packed crystal structures. In face-centered cubic (fcc) materials, they play a significant role in governing deformation behavior. When the SF energy is sufficiently low, these planar defects form by the dissociation of dislocations [9], or in larger quantities through mechanisms such as transformation-induced or twinning-induced plasticity [10, 11]. Once present, SFs can act as barriers to further dislocation motion [12, 13]. At elevated temperatures, they can also undergo local compositional changes due to atomic-scale Suzuki segregation along the fault plane [14, 15]. Thus, SFs represent a critical microstructural feature in many fcc-based alloys.

SF-based deformation mechanisms have received particular attention in $\gamma/\gamma'$-strengthened superalloys with the base element Ni or Co (e.g. [16-27]), which are tailored to possess excellent mechanical properties at temperatures near their melting points [28, 29]. In this class of alloys, the $\gamma$ matrix phase adopts the fcc structure; coherently embedded into this phase are cuboidal precipitates of the $L1_2$-ordered $\gamma'$ intermetallic phase, which is typically based on a composition of $Ni_3Al$ for Ni-base and $Co_3(Al,W)$ for Co-base superalloys. The interplay between the two phases leads to precipitation strengthening, as dislocations encounter a barrier to entering the precipitates: Perfect $^a/_2\langle 110\rangle$ dislocations in the γ phase become superpartials in the γ′ phase, where they would create a high-energy antiphase boundary [30]. Furthermore, perfect $a\langle 110\rangle$ superdislocations in γ′ are energetically unfavorable due to their large Burgers vectors. Thus, dislocation configurations entering the γ′ phase often dissociate into partial and superpartial dislocations, leaving behind planar faults including SFs [16, 22, 31]. Besides shearing, SFs in superalloys have also been observed to form through dislocation climb [32]. Depending on alloy composition and deformation parameters, SF-based deformation mechanisms can be responsible for a majority of the plastic deformation in these alloys [33, 34].

SFs also play an important role in semiconductors, as they locally modify the band structure [35, 36] and can interact with intrinsic point defects [37], or become decorated by impurities [38, 39]. As a consequence, SFs can severely affect electronic and opto-electronic device performance [40, 41], but may also give rise to intriguing opto-electronic properties, such as ultra-narrow quantum well emission [42].

For the investigation of crystal defects, (scanning) transmission electron microscopy ((S)TEM) is particularly suitable. A schematic overview of different TEM-based analysis methods to determine the type of a given SF, i.e., whether it is intrinsic (ISF) or extrinsic (ESF), is shown in Fig. 1.

Methods for determining the types of planar faults in TEM based on fringe contrast (Fig. 1a) have been available for many decades and continue to be applied today [34, 43-45]. In fcc materials, as fringe contrast requires an inclined SF orientation, [001] foils are typically used, where all four {111} glide planes are inclined (Fig. 1d). For instance, the method by Gevers et al. [45] requires a dark-field micrograph recorded under a two-beam condition. The origin of the **g** vector is placed at the center of the fault, and the fault type is determined based on the nature of the reflection and whether the **g** vector points toward or away from the bright outer fringe. Since fringe contrast is highly sensitive to small changes in excitation error, it can be challenging to obtain reliable results using these fringe-based methods, especially when the crystal is strongly bent. Additionally, SF analysis becomes difficult when the fault segment is very short or when two or more SFs overlap.

Continuous improvements in high-resolution TEM (HRTEM) and high-resolution STEM (HRSTEM) have enabled more robust analysis of stacking faults (SFs) through the direct interpretation of atomic-scale stacking sequences in micrographs (Fig. 1b) (e.g. [20, 46] for superalloys). This requires the SF to be in edge-on orientation; therefore, in fcc materials, [110] foils are commonly used, where two of the four {111} glide planes are edge-on (Fig. 1d). In the $L1_2$ structure, given sufficient superlattice contrast of the atomic columns, it is also possible to determine reliably whether a fault is of complex or superlattice nature [47]. A major limitation of this method is the access to only two of the four glide planes in [110] foils, as the other two are at an inclined angle to the viewing direction, and the complete inadequacy of [001] foils, where all four glide planes are inclined. Schematic drawings of the orientations of the {111} SF habit planes in fcc crystals viewed along the [110] and [001] zone axes (ZAs) are shown in Fig. 1d.

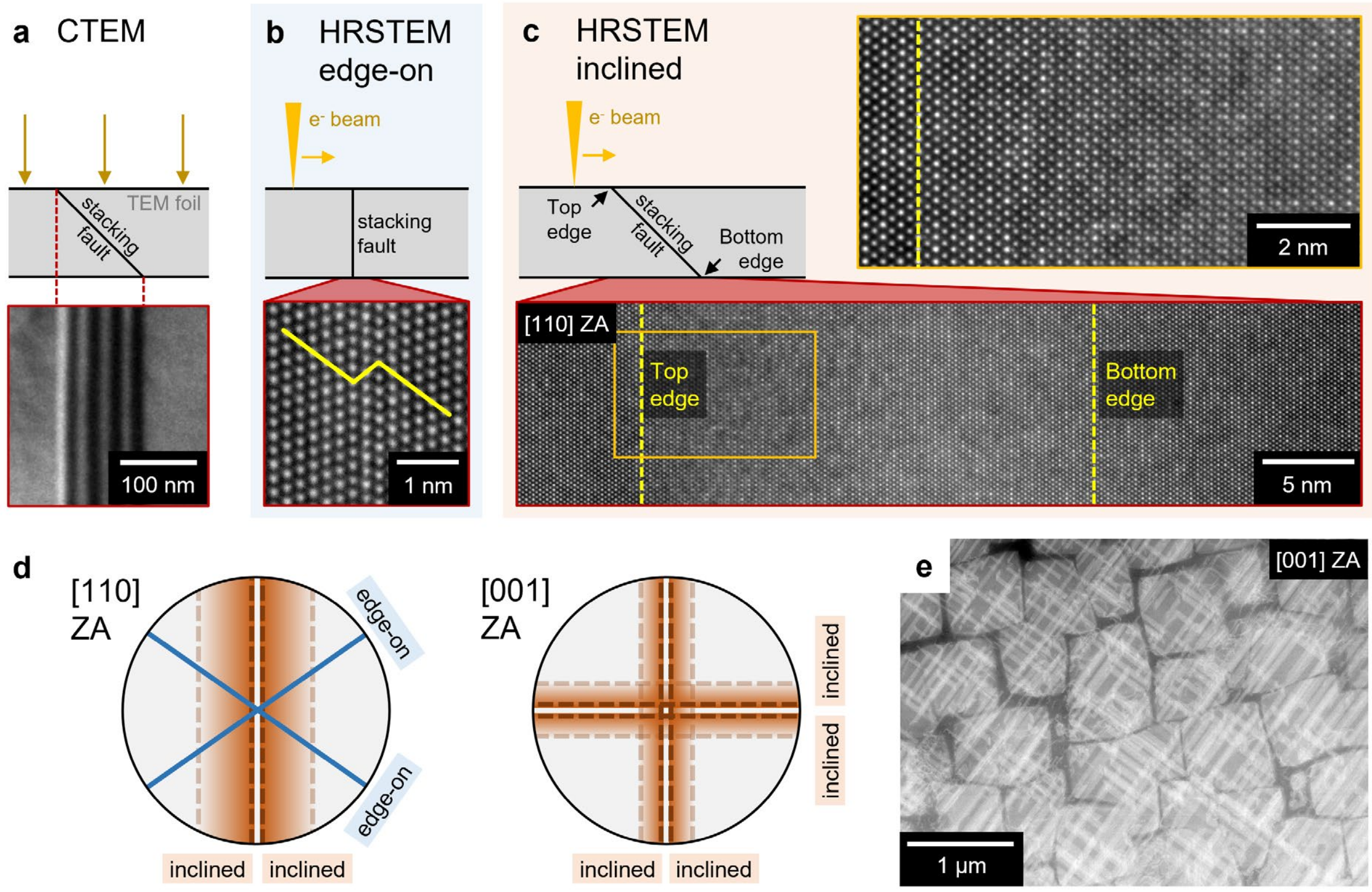

Fig. 1: Schematic overview of different SF analysis methods in the TEM, and crystallographic orientation of SFs in a TEM foil. (a) Conventional fringe-based contrast at inclined SFs. (b) HRSTEM-based method observing the stacking sequence edge-on. (c) Novel method enabling the HRSTEM-based analysis of inclined SFs. The SF is imaged in the region where it intersects the upper surface of the TEM lamella (top edge). In the example, this occurs on the left side of the image; toward the right, the depth of the SF gradually increases, as shown in the schematic side view. The magnified HRSTEM image of the SF near the top edge reveals a superposition of the lattices above and below the fault, enabling a direct evaluation of the projected lattice shift. (d) Schematic representation of the edge-on or inclined orientations of the {111} habit planes of SFs when the crystal is viewed in a [110] or [001] ZA. (e) ADF-STEM micrograph of the microstructure of the ERBOCo-4 alloy after deformation, exhibiting a high SF density.

To overcome these limitations, we present a new method in this work enabling the analysis of SFs on inclined planes using HRSTEM (Fig. 1c), combining the advantages of this imaging technique with access to all four glide planes in both [001] and [110] zone axes (ZAs) when used in combination with edge-on analysis. Projected structures of the different fault types are investigated in the fcc and $L1_2$ structures in both [001] and [110] projection and compared to experimental HRSTEM micrographs acquired in a CoNi-base superalloy as well as corresponding multislice simulations. The HRSTEM contrast observed at inclined SFs is rationalized by probe intensity distribution simulations. Furthermore, the method is successfully applied in a state-of-the-art oxide-dispersion strengthened alloy as well as a semiconductor with sphalerite structure, and it is demonstrated that the analysis in inclined geometry can be expanded to the bounding partials of SFs to reveal their formation mechanisms.

## Results and Discussion

The analysis method presented here is based on the fact that an SF in inclined orientation gives rise to additional bright spots in a high-angle annular dark-field (HAADF) HRSTEM micrograph. An example from a relatively thin sample area is shown in Fig. 1c. This micrograph contains both edges of the SF where the fault is confined by the foil surfaces as indicated in the schematic side view; in the micrograph these edges are oriented vertically. The atomic columns in both half-crystals on either side of the SF do not align in projection; therefore, a superposition of both lattices is visible where the half-crystals overlap, and the observed shift between them depends on the SF type as will be shown below.

The micrograph at the top right of Fig. 1c shows a magnified view of the region where both lattices are visible. Notably, this region of visible overlap does not extend across the entire projected width of the SF, but it is confined to one of the edges. This is because, as soon as the top half-crystal has reached a sufficient thickness within the foil, the probe can no longer establish a substantial enough channeling condition necessary for atomic-column HAADF contrast [48, 49] in the lower half-crystal. Since this interpretation of the images is not obvious, and has not been considered previously, the details of HAADF contrast formation at an inclined SF are elucidated in greater detail in Section 2.4. Consequently, the superposition of both lattices is only visible where the SF is near the top of the foil, making the top edge of the SF easily distinguishable from the bottom edge. As demonstrated in Fig. 1e, SFs themselves can easily be discerned in ADF-STEM micrographs by the bright band of contrast they produce.

This micrograph also shows the single crystalline, two-phase microstructure of the ERBOCo-4 alloy examined in this study, consisting of cuboidal precipitates of the $\gamma'$ phase with $L1_2$ crystal structure embedded into the fcc $\gamma$ matrix phase.

In the following, to facilitate a systematic comparison between projected intrinsic and extrinsic fault structures, SF images will be oriented so that the top edge is on the left and the depth of the SF in the foil increases towards the right. First, expected projected fault structures in the fcc and $L1_2$ structures are discussed, followed by an analysis of faults using high-resolution micrographs.

## Projected fault structures in fcc and $L1_2$

Fig. 2a visualizes the translation vectors and projected structures of ISFs (green) and ESFs (purple) on the inclined $(111)$ plane in the fcc structure for the [110] and [001] ZAs. Due to crystal symmetry, this discussion also applies to other $\langle 110\rangle$- and $\langle 001\rangle$-type ZAs combined with other inclined $\{111\}$ glide planes, respectively, as long as the fault is oriented so that its top edge is on the left and its depth within the foil increases towards the right. As indicated in the illustration, the angle of the $(111)$ plane to the foil plane differs between the two ZAs (approx. 35.3° for the [110] ZA and 54.7° for the [001] ZA).

In the side view of the foil (see upper part of Fig. 2a), the left half crystal will be referred to as the reference lattice (atoms represented by circles, solid unit cell), and the right half crystal as the shifted lattice (atoms represented by squares, dashed unit cell). Therefore, for this analysis, it is more intuitive to describe translations as being applied to the top crystal (contrary to the usual convention for the fringe contrast method where translations are applied to the bottom crystal [50]). The lattice translations $\mathbf{R}_{\mathrm{ISF}}$ and $\mathbf{R}_{\mathrm{ESF}}$ introduced by the two fault types are exemplarily shown as arrows. It should be noted that, in each glide plane, the lattice translation required for an ISF can be achieved by three distinct ${}^{a}/_{6}\,\langle 112\rangle$-type translation vectors with different directions; the same is true for the ${}^{a}/_{3}\,\langle 112\rangle$-type net translation vectors of ESFs. However, in both cases, they result in identical fault structures; therefore, only one of each is described without loss of generality. A detailed description of translation vectors and their projections [9, 51] can be found in Supplementary Notes 1, 2, and Supplementary Fig. 1 in the Supplementary Information (SI).

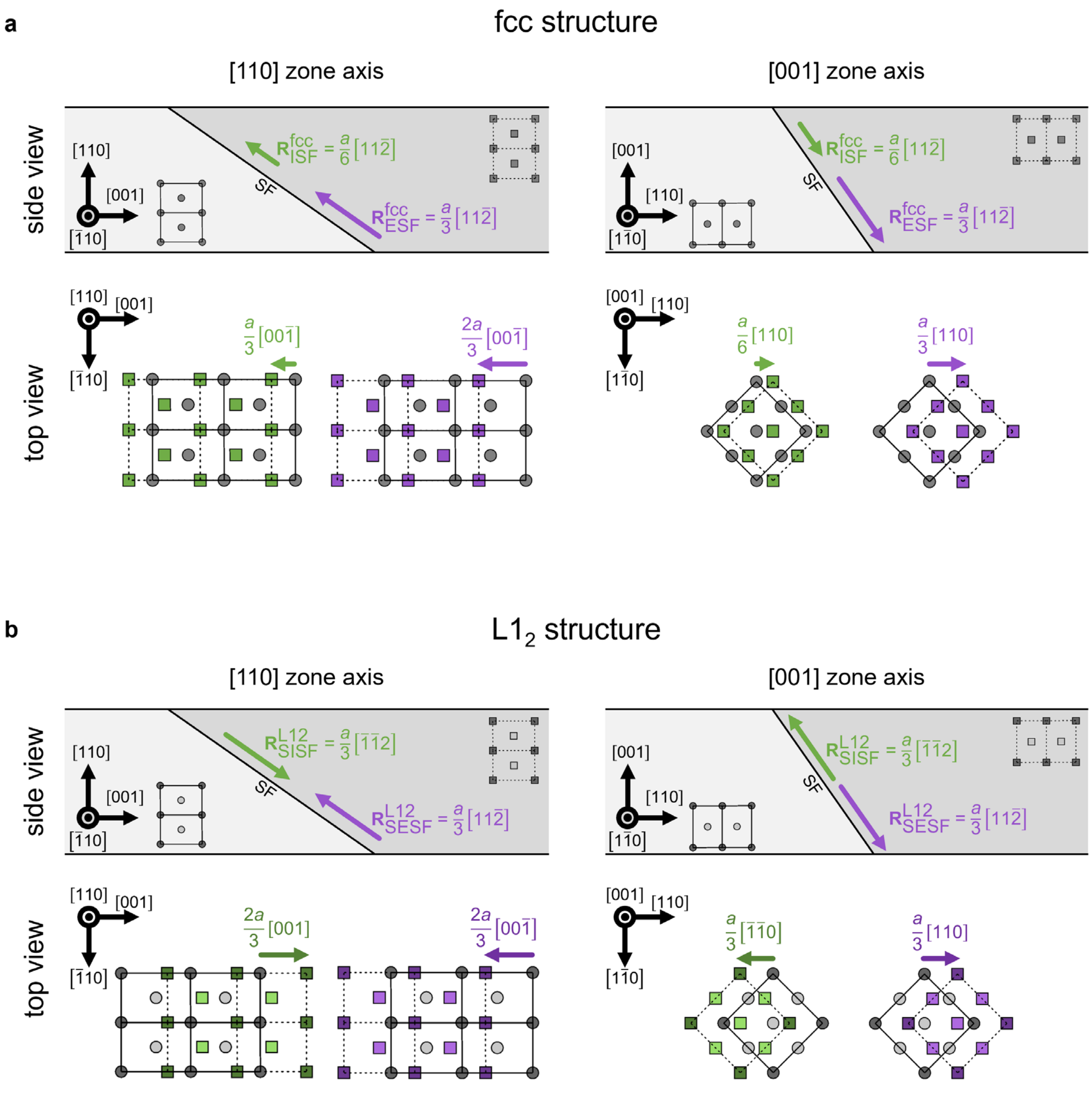

Fig. 2: Schematic overview of inclined SFs, their translation vectors **R**, and their projected structures in the [110] and [001] ZAs in the (a) fcc and (b) $L1_2$ crystal structure. The atoms of the reference lattice are represented by circles; those of the shifted lattice are represented by squares. The latter are colored green for (S)ISFs and purple for (S)ESFs in the top-view superposition of both lattices. Brighter and darker colors represent the two different sublattices of the ordered structure.

The projected fault structures, i.e., the superpositions of reference and shifted lattice as observed in the respective ZA, are visualized in the lower part of Fig. 2a. In the [110] ZA, the lattice translation introduced by the ISF leads to a projected shift along the [001] direction to the left (towards the fault's top edge) by one third of the lattice periodicity ($^{a}/_{3}$ $[00\bar{1}]$). The atomic columns of the shifted lattice can therefore be described as appearing to the left of those of the reference lattice. On the other hand, the ESF lattice translation is twice as large, with a projected shift $^{2a}/_{3}\,[00\bar{1}]$. Effectively, the atomic columns of the shifted lattice appear to the right of those of the reference lattice. (It should be noted that an ESF, which consists of two ISFs on neighboring planes, possesses a central fault plane which is not part of either half crystal. However, as this plane is only a single atom thick, its contribution to HAADF image contrast is negligible; therefore, it is sufficient to consider only the net translation of the ESF here.)

In the [001] ZA, the projected shifts corresponding to $R_{ISF}$ and $R_{ESF}$ appear along the [110] direction; their values are $^{a}/_{6}\,[110]$ and $^{a}/_{3}\,[110]$, corresponding to a shift to the right by one third and two thirds of the lattice periodicity, respectively. Thus, for an ISF, columns of the shifted lattice appear to the right, whereas for an ESF, they appear to the left in this ZA.

In the $L1_2$ structure, further distinctions between fault types – complex or superlattice SFs – arise from the presence of superlattice ordering (cf. Supplementary Fig. 1 and Supplementary Note 1 in the SI). However, regarding the difference between intrinsic and extrinsic SFs, the analysis of inclined SFs in the $L1_2$ structure is fundamentally very similar to that in the fcc structure as only the stacking sequence, i.e., the position of atomic columns, needs to be considered, as opposed to which columns appear brighter and which appear darker. Typically, superlattice SFs have considerably lower energies than their complex counterparts [30, 51]; thus, extended faults can be assumed to be non-complex. The analysis method presented here is therefore aimed at distinguishing between superlattice intrinsic (SISFs) and superlattice extrinsic stacking faults (SESFs). A discussion for inclined complex SFs and a comparison to superlattice SFs can be found in Supplementary Note 3 and Supplementary Fig. 2.

Fig. 2b gives a schematic overview of the translation vectors and projected structures of SISFs and SESFs in the $L1_2$ structure on the inclined (111) plane for the [110] and [001] ZAs. As above, this discussion is also valid for other inclined {111} glide planes viewed in other ⟨110⟩- and ⟨001⟩-type ZAs, respectively. Again, each fault is oriented so that its top edge is on the left and its depth in the foil increases towards the right.

In Fig. 2b, atoms of the reference and shifted lattice are represented again by circles and squares, respectively. To differentiate between the two sublattices of the $A_3B$-type $L1_2$ structure, brighter (A) and darker shades (B) of the respective colors are used. As schematically illustrated in the side view, the translation vector associated with a SISF in the $L1_2$ structure is twice as long as that of an ISF in the fcc structure, and its direction is reversed. Its reversed direction makes the SISF distinguishable from the SESF, whose net translation vector has the same magnitude (see Supplementary Note 1 and Supplementary Fig. 1 for details). As above, it should be noted that, for each fault type, different translation directions are available; however, they result in identical fault structures, and so only one is shown.

From the visualization of projected fault structures (lower part of Fig. 2b), it is evident that inclined SISFs and SESFs can be distinguished in the same way as ISFs and ESFs above: In [110] projection, for the SISF, the shifted lattice is translated to the right (along the [001] direction) by two thirds of the lattice periodicity, its atomic columns therefore appear to the left of the reference lattice, and vice versa for the SESF. In [001] projection, the translation appears along the [110] direction, and the situation is reversed.

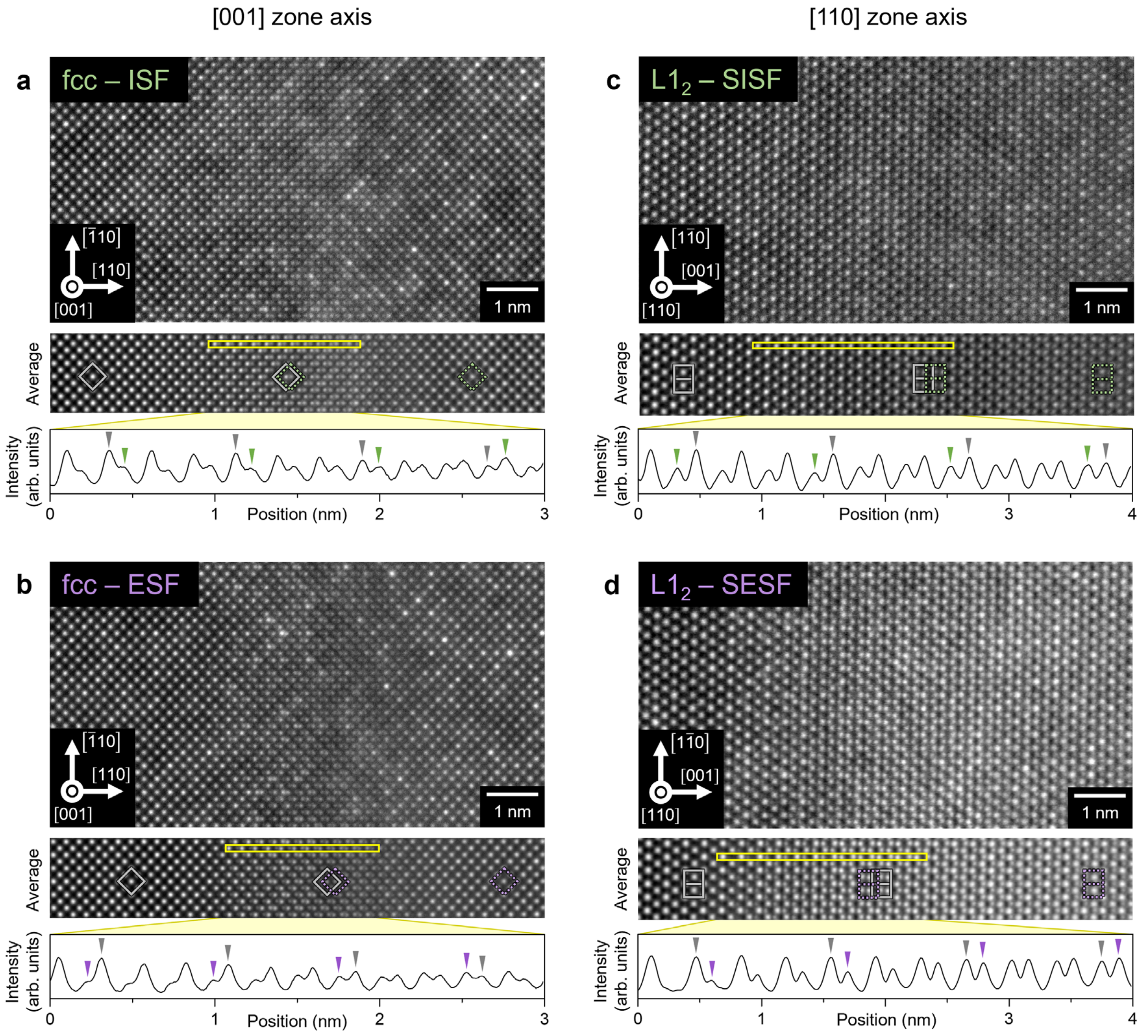

Fig. 3: Experimental HRSTEM micrographs of SFs in the (a,b) fcc and (c,d) $L1_2$ structures. (a) ISF and (b) ESF in the fcc structure recorded in the [001] ZA as well as (c) SISF and (d) SESF in the $L1_2$ structure recorded in the [110] ZA are shown. Below each micrograph, averages of periodically repeating units are shown along with line scans extracted from the regions marked by yellow rectangles. In the intensity profiles, some peaks of the reference and shifted lattice are highlighted by grey and colored triangles, respectively. The positions of unit cells of the reference and shifted lattice are indicated in the averaged images.

Overall, a distinction between (S)ISFs and (S)ESFs is therefore possible in both ZAs, which will be demonstrated by experimental examples in the following (Fig. 3).

Fig. 3a,b presents experimental HRSTEM micrographs in the [001] ZA recorded at the top edges of SFs in fcc γ channels of the microstructure. (Surface roughness of the TEM foil at the atomic scale may lead to the top edge of an SF not appearing as a perfectly straight line at this magnification.) The two cases described above, where a new lattice appears shifted along the [110] direction to the right (Fig. 3a) or left (Fig. 3b) of the reference lattice, can be clearly distinguished. The difference between the two is made even clearer by exploiting the fact that the micrographs consist of periodically repeating units in the vertical direction and averaging these periodic units as shown below the micrographs. Intensity line scans measured horizontally across the row of spots marked by yellow rectangles in the averaged experimental data also readily visualize the difference between the appearance of columns of the shifted lattice on the right and on the left side of the reference lattice, respectively, allowing for the distinction between ISF and ESF in this ZA.

In the examples shown for the $L1_2$ structure in the [110] ZA (Fig. 3c,d), the appearance of columns of the shifted lattice to the left (Fig. 3c) or to the right (Fig. 3d) of the reference lattice can be clearly distinguished, revealing that the former is an SISF and the latter an SESF. It should be noted that the superstructure gives rise to two different types of atomic columns in [110] projection: those containing only atoms of the A sublattice, and mixed columns containing both. Due to the difference in average atomic number between these two types of columns, they exhibit a slight difference in intensity (*Z*-contrast).

To obtain further confirmation for the correspondence between fault structure and resulting image contrast, HAADF-STEM multislice image simulations were performed for the intrinsic and extrinsic faults in the fcc and $L1_2$ structures in both ZAs (see Supplementary Fig. 3). As a final experimental validation, analyses based on three different methods – HRSTEM, conventional fringe contrast [45], and convergent-beam electron diffraction (CBED) [52-54] – were performed all on the same SF, with all results in agreement (see Supplementary Fig. 4).

In summary, the ability to distinguish between inclined intrinsic and extrinsic SFs in the fcc and $L1_2$ structures has been demonstrated for the [110] as well as [001] ZAs. In the following sections, we investigate the mechanism of contrast formation, discuss practical aspects of the technique, and present several applications demonstrating its broad applicability.

## Contrast formation at inclined faults

To gain a deeper understanding of the contrast formation where reference and shifted lattice overlap in the image, multislice probe propagation simulations have been carried out (Fig. 4).

As a reference, Fig. 4a shows a simulated HAADF-STEM micrograph of an inclined ISF in the fcc structure. Three regions are highlighted: the perfect crystal on the left, the region where the SF is near the top surface and a superposition of reference and shifted lattice is visible in the HAADF-STEM image, and the region to the right where the SF is deeper within the foil and only the shifted lattice is visible in the micrograph. The contrast formation for each of these three situations is discussed in the following.

Fig. 4b shows the probe intensity distribution projected along the viewing direction through the simulated cell of a perfect Ni crystal illuminated in the [001] ZA. The probe is centered on an atomic column, as indicated by the yellow cross in the top-view inset at the upper left. A corresponding side view of a portion of the crystal, including the probe position, is shown below. The probe intensity distribution on the right clearly reveals strong channeling of the probe along the atomic column, consistent with the expected behavior in a perfect crystal [55]. Channeling plays a significant role in HAADF contrast formation, as the image intensity primarily arises from electrons scattered out of the strongly channeled probe [56]. Thus, as illustrated on the left side of Fig. 4e, scattering from the probe at the atomic column contributes to the HAADF-STEM image intensity down to a considerable depth within the crystal, although the probe – and consequently the high-angle scattering – gradually decreases in intensity with increasing depth.

For the simulations shown in Fig. 4c, an ISF has been introduced into the crystal at a depth of 4.5 nm. In the schematic illustration, the upper, shifted portion of the lattice is represented by green atoms. As indicated in both the top and side views, probe propagation simulations were carried out for two distinct probe positions: one centered above an atomic column in the lower, reference lattice (grey cross), and the other above a column of the upper, shifted lattice (green cross).

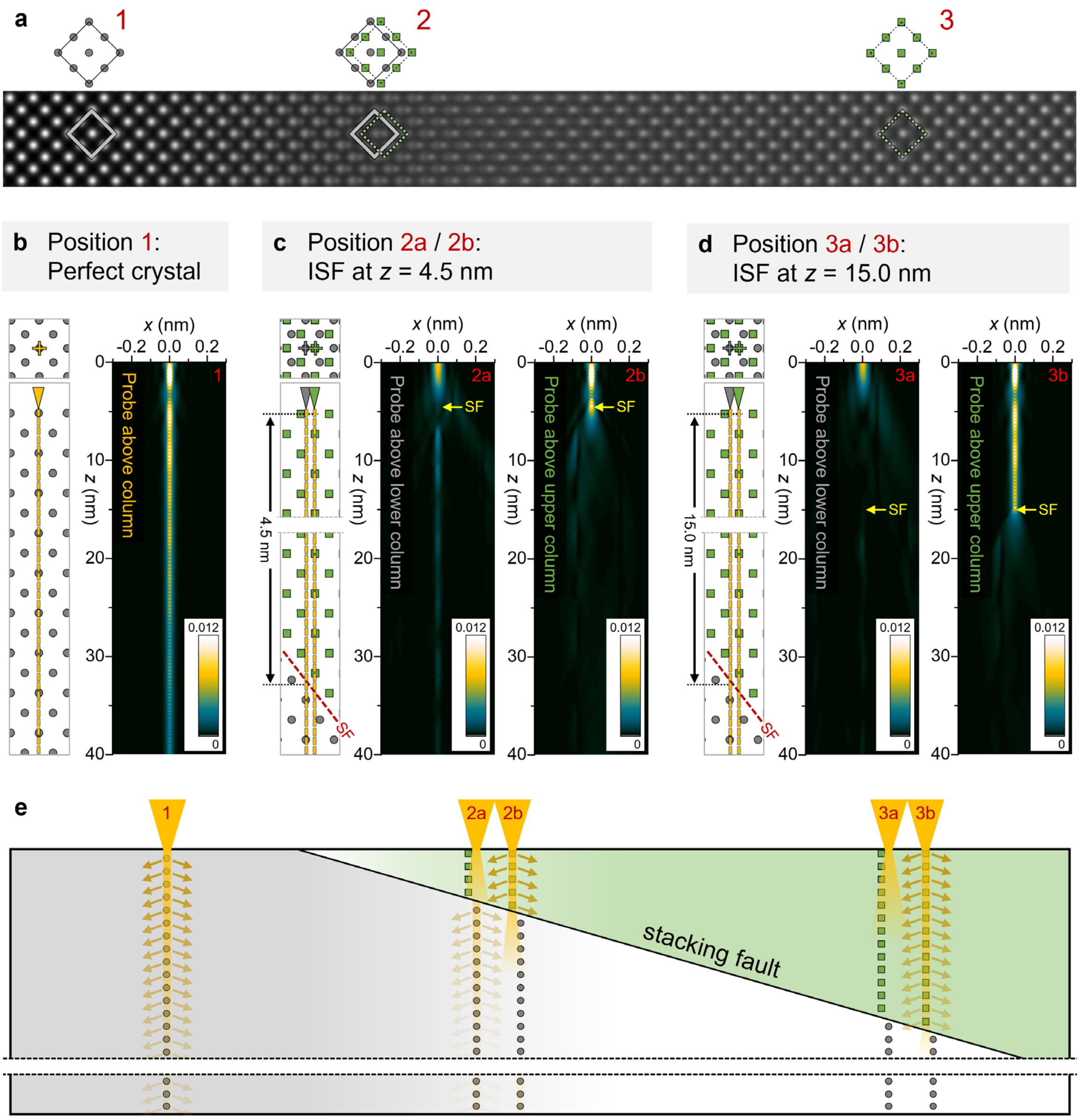

Fig. 4: Simulated probe intensity distributions and HAADF-STEM contrast formation at inclined SFs. (a) HAADF-STEM multislice image simulation of an inclined ISF in fcc Ni in the [001] ZA at a foil thickness of ≈40 nm. Three regions are highlighted: perfect crystal (1), inclined SF near the top surface with visible superposition of reference and shifted lattice (2), and inclined SF at larger depth with only the shifted lattice visible (3). (b-d) Probe propagation simulations in three different structures: (b) perfect fcc crystal, (c) crystal with an inclined ISF at a depth of 4.5 nm, (d) crystal with an inclined ISF at a depth of 15.0 nm. The left side of each subfigure shows a schematic visualization of part of the corresponding crystal in top view (top) and side view (bottom) along with the chosen probe positions. In (b), the probe (yellow) is centered above an atomic column; in (c) and (d), two cases are shown where the probe is centered above a column of the crystal below the SF (grey) and above SF (green), respectively. The plots show a vertical slice of the probe intensity distribution (lateral position *x*, depth *z*) at the location of the incident probe. To enhance the visibility of lower intensities, a non-linear scale is used ($\gamma$=0.58); plots in a linear and a logarithmic scale can be found in Supplementary Fig. 5. In all cases, the size of the simulated cell is ≈5 x 5 x 40 nm$^3$. (e) Schematic side-view illustration of a foil containing an inclined SF, visualizing where channeling of the electron probe (depicted in yellow) occurs, which leads to high-angle scattering (represented by arrows) and a contribution to HAADF signal.

When the probe is centered on a column of the lower lattice, it initially traverses the upper portion of the sample in a non-channeling condition, resulting in significant delocalization of its intensity. However, as the residual probe propagates into the lower lattice, it still enters a channeling condition, albeit at reduced intensity. This behavior is illustrated in Fig. 4e and explains why, up to a certain thickness of the upper lattice, atomic columns of the lattice below remain visible, leading to the characteristic region of overlapping lattices in HRSTEM micrographs.

On the other hand, when the probe is centered on a column of the upper lattice, strong channeling occurs until the probe encounters the lower lattice, where the channeling condition is disrupted. As a result, only the region above the stacking fault contributes significantly to the HAADF atomic-column contrast, while the region below contributes minimally. This is illustrated in Fig. 4e and explains why atomic columns of the upper lattice remain clearly visible in micrographs, even when these columns are relatively short.

Finally, in Fig. 4d, the ISF is positioned deeper within the crystal, at a depth of 15.0 nm. As before, probe propagation simulations were performed each with the probe placed above an atomic column in the lower lattice (grey cross) and one in the upper, shifted lattice (green cross).

When the probe is now placed above a column of the lower lattice, it remains in a non-channeling condition over a longer distance compared to the previous case, leading to much greater delocalization of its intensity. By the time the SF is reached, virtually no probe intensity remains to establish a channeling condition. Consequently, there is no significant scattering of electrons to large angles from this column, and the lower lattice does not produce observable HAADF contrast.

In contrast, a probe centered on a column of the upper lattice immediately enters a channeling condition, which it maintains over a long propagation distance until it reaches the lower lattice where it becomes dechanneled. Thus, atomic-column HAADF contrast is generated from the upper crystal.

In summary, going from left to right in Fig. 4a and e, as the SF appears and its depth increases, the atomic columns of the shifted lattice become longer, allowing the probe to maintain channeling along them over a greater distance. This leads to an increasing HAADF contrast contribution from the upper, shifted lattice. Simultaneously, channeling at the atomic columns of the lower reference lattice progressively diminishes as the probe undergoes increasing dechanneling within the upper crystal before reaching these columns, thereby reducing their contribution to the HAADF contrast. In this way, a region of overlapping contrast from both lattices emerges in the HAADF signal only when the SF is near the top surface.

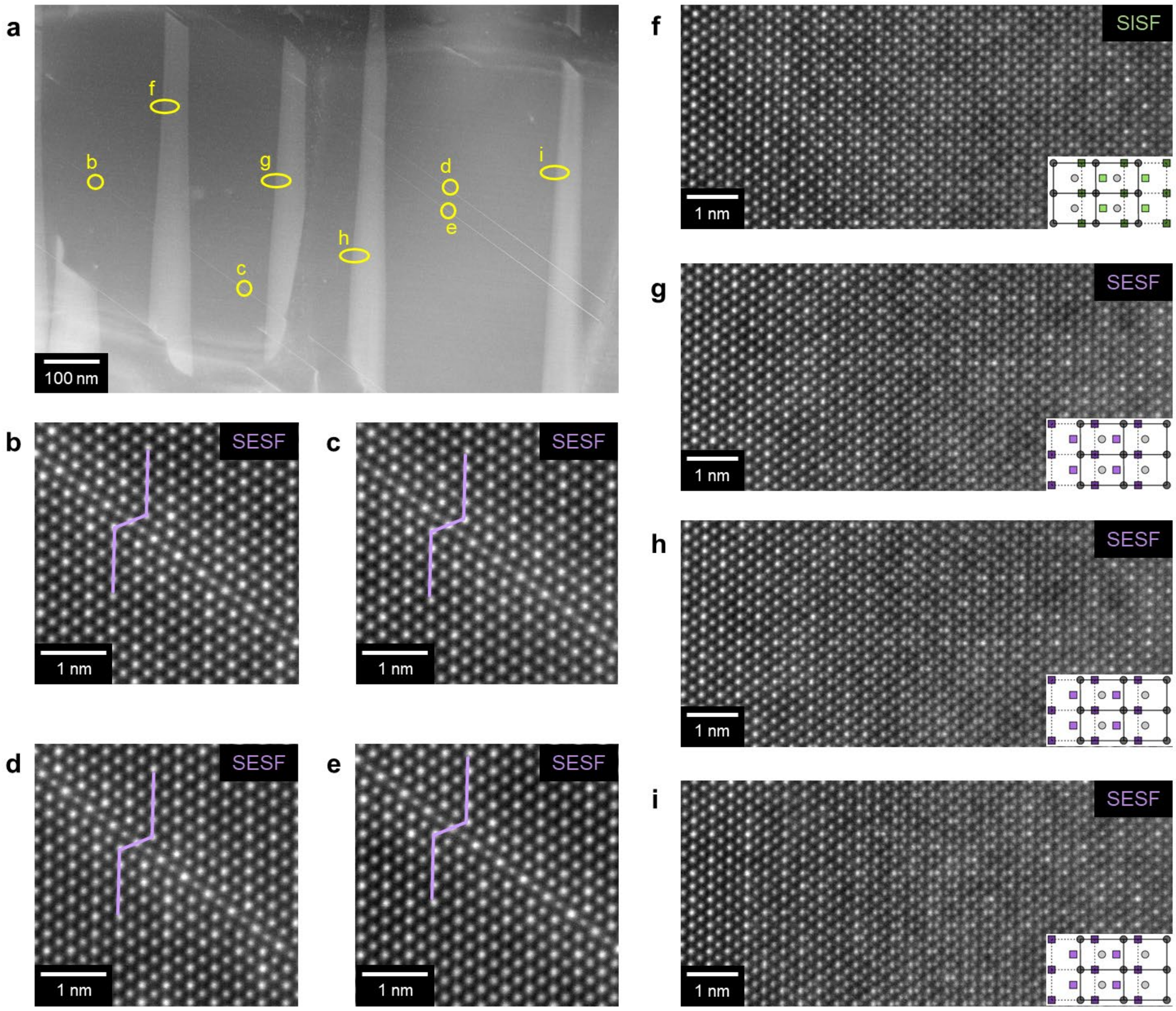

Fig. 5: Analysis of edge-on and inclined SFs in the [110] ZA at the same time. (a) ADF-STEM micrograph of the microstructure containing SFs on different glide planes. HRSTEM micrographs of faults on (b-e) edge-on and (f-i) inclined glide planes, from the locations marked in (a), revealing each fault's type.

## Advantages and practical considerations

So far, SFs could only be investigated either edge-on by HRSTEM/HRTEM or in inclined orientations by conventional TEM techniques. A key advantage of the present method is that it enables the analysis of both geometries using HRSTEM alone. This is demonstrated in Fig. 5 for the [110] projection, where SFs in both geometries coexist. Within the region shown in Fig. 5a, locations at eight SFs are marked where HRSTEM micrographs were recorded. The corresponding fault types are readily and reliably identified for both SFs on edge-on (Fig. 5b-

e) and inclined (Fig. 5f-i) glide planes. In the [001] projection, the impact of the present method is even greater, as all SFs are inclined (Fig. 1d) and classical edge-on HRSTEM/HRTEM analysis is therefore entirely inapplicable.

Furthermore, HRSTEM analysis of inclined SFs has additional advantages over the classical fringe-based method. It is reliable even at very low thicknesses – and, due to the local nature of HRSTEM contrast, for very short SF segments – where fringe contrast may not be useful. At the same time, the method we have presented is still viable even at a relatively high foil thickness. Two examples are presented in Fig. 6: one in the [110] ZA at a thickness of ≈100 nm (Fig. 6a) and one at a thickness of ≈130 nm in the [001] ZA (Fig. 6b). The foil thickness in the respective regions was calculated from the projected thickness of SFs in lower-magnification images as well as the known relationship between fault plane and foil normal [34]. Even at these relatively high foil thicknesses, the HRSTEM contrast can be clearly interpreted with respect to the fault type. Sample thicknesses of ≈100 nm are routinely achieved by various preparation methods across different material systems, making this SF analysis method readily accessible from a sample-preparation perspective. In this context, we note that for SFs in the semiconductor GaP, successful analysis was achieved even at sample thicknesses of ≈150 nm (see below) and ≈270 nm (Supplementary Fig. 11f,g).

Another advantage, especially at higher SF densities, is the ability to analyze SFs which overlap in projection, as only the contrast near the top edge of the fault is relevant to the analysis. In the example shown in Fig. 6c in the [001] ZA, the top edges of the two SFs towards the left side are separated by only 6 nm (see inset). The HRSTEM micrograph in Fig. 6d contains the top edges of both SFs, and even at this small separation, each SF can be analyzed separately: the fault on the left is intrinsic, whereas the fault on the right is extrinsic. Another example of overlapping SF analysis in the [110] ZA is shown in Supplementary Fig. 6, along with a comparison to fringe-based contrast, which becomes complicated and therefore can hardly be used for SF analysis under these conditions.

More generally, the robustness of the proposed HRSTEM method for SF analysis benefits from its highly localized evaluation of the image contrast. This advantage is not limited to the analysis of overlapping SFs but also extends to specimens exhibiting local foil bending, a common phenomenon in TEM samples and therefore of considerable practical relevance. Since only a small region (typically less than 10 nm) adjacent to the intersection of the SF with the specimen surface needs to be aligned to the desired zone axis, moderate local orientation variations have little influence on the analysis. In contrast, conventional CTEM approaches based on SF fringe contrast require suitable diffraction conditions to be maintained over much larger regions, typically on the order of several tens to hundreds of nanometers. Local foil bending can therefore lead to significant variations in the diffraction conditions across the analyzed area, resulting in corresponding changes in the fringe contrast and making the interpretation considerably less reliable.

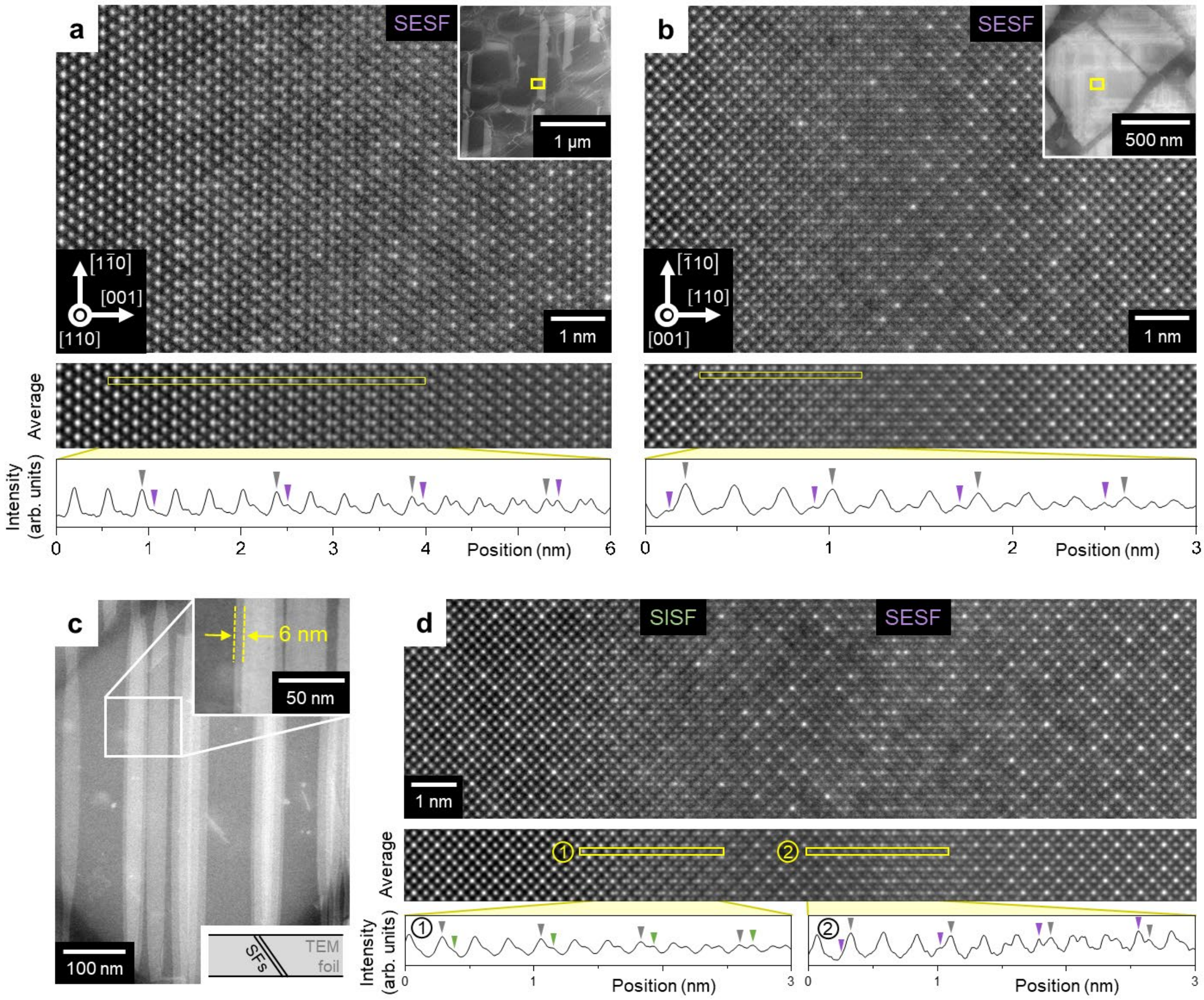

Fig. 6: Analysis of inclined SFs at high foil thickness, as well as faults overlapping in projection. (a,b) HRSTEM micrographs of SFs in the $L1_2$ structure (a) in the [110] ZA at a thickness of ≈100 nm and (b) in the [001] ZA at a thickness of ≈130 nm. The insets show ADF-STEM micrographs of the sample regions where the respective HRSTEM micrograph was recorded (yellow rectangle). The local foil thickness was calculated geometrically from the projected width of the SF as well as its inclination angle. (c) ADF-STEM micrograph of closely overlapping SFs in the [001] ZA whose top edges are separated by only 6 nm (see magnified inset). (d) HRSTEM micrograph containing the top edges of both SFs, from which both faults can be characterized. Below each HRSTEM micrograph, the periodically averaged signal is shown along with intensity linescans from the areas marked by yellow rectangles. In the intensity profiles, some peaks of the reference and shifted lattice are highlighted by grey and colored triangles, respectively.

Given the inclined geometry of the SF in the foil, the location where the region of visible overlap appears in high-resolution micrographs varies slightly with defocus (see Supplementary Fig. 7 for an experimentally recorded defocus series). The highest resolution, especially in the superposition region, is achieved with the focal plane near the top surface of the foil. Furthermore, the width of this region depends on the effective depth of focus, and thus on the beam convergence angle (see Supplementary Fig. 8 for a direct comparison). For the highest resolution, the optimal convergence angle for the specific microscope and correction optics should be used. Both effects have been reproduced in multislice image simulations (Supplementary Fig. 9).

It is known that the presence of an amorphous surface layer on the foil, which may result from specimen preparation, can reduce HRSTEM contrast [57]. To assess the effect of specimen preparation on high-resolution contrast at inclined SFs, HRSTEM micrographs from three foils prepared using different procedures—electropolishing with and without subsequent ion polishing, and focused ion-beam liftout—were compared. All three procedures proved suitable for resolving individual atomic columns, even in the overlap region between reference and shifted lattice, thereby enabling reliable identification of the SF type (see Supplementary Fig. 10).

Generally, the resolution requirement to resolve atomic columns of both lattices in the overlapping region individually is rather high, as exemplified by the projected distance between the nearest atomic columns of the reference and shifted lattice. In the [110] ZA, this distance equals $^{a}/_{3}$ [001] which, in the case of fcc Ni ($a$ = 3.52 Å), corresponds to 117 pm. In the [001] ZA, the projected distance between columns is even smaller, at $^{a}/_{6}$ [110], corresponding to 83 pm. To achieve this resolution, careful aberration correction is crucial. However, the method presented in this work requires only the direction of relative shift between both lattices to be detected (e.g., by geometric phase analysis; see below). The resolution requirement for this distinction is considerably lower, making the application of this method possible even in non-aberration-corrected STEM.

A limitation of the method, inherent to the inclined SF geometry and therefore also shared with fringe-based methods, is that only the net translation vector of a fault can be extracted. Consequently, if microtwins or configurations consisting of closely spaced SFs (e.g., [58, 59]) are present in the microstructure, distinguishing them from intrinsic or extrinsic SFs becomes difficult without considering the bounding partial dislocations. Only high-resolution edge-on analysis in the [110] ZA allows the exact local stacking sequence of all planes within the fault to be resolved.

## Applications of stacking fault analysis

Besides superalloys, the method presented above was successfully applied to other materials: a state-of-the-art oxide-dispersion strengthened alloy, as well as a semiconductor. Furthermore, the high-resolution analysis of inclined faults has been extended to the characterization of their bounding partial dislocations. These results are presented in the following.

GRX-810 (Fig. 7) is an oxide-dispersion strengthened (ODS), additively manufactured (AM) CoCrNi-based medium entropy alloy [60]. Using resonant acoustic mixing (RAM), a nanoscale yttria coating is applied to the base alloy powder, which is subsequently printed via Laser Powder Bed Fusion (L-PBF) [60, 61]. The resulting alloy features a fine, homogeneous oxide dispersion and exhibits exceptional high-temperature performance, including >1,000-fold improvements in creep resistance relative to widely used printable high-temperature alloys under comparable conditions [60]. This behavior is attributed to the combined effect of nanoscale oxides, grain boundary-stabilizing carbides, and the system's low stacking fault energy [62, 63].

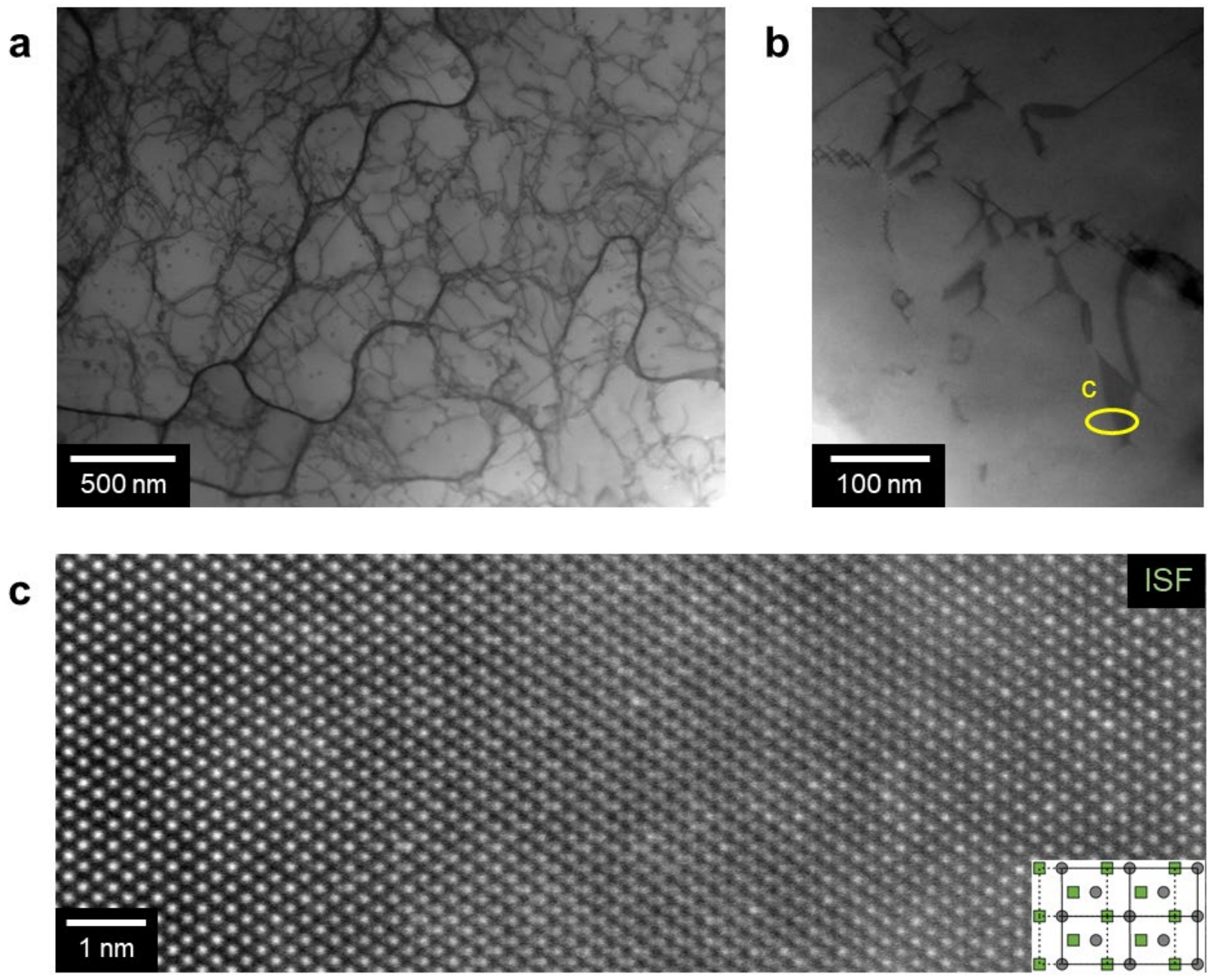

Fig. 7: Analysis of an inclined SF in alloy GRX-810. (a) BF-STEM micrograph of the microstructure. (b) BF-STEM micrograph showing SF nodes. (c) HRSTEM micrograph of the SF marked in (b), revealing it to be intrinsic.

As shown in Fig. 7a, the microstructure of GRX-810 consists of a fine subgrain structure with a high dislocation density stabilized by oxide particles. These dislocations are dissociated, forming stacking fault node structures. Such a structure is shown in Fig. 7b; the inclined SF in the highlighted location is clearly revealed to be intrinsic from the high-resolution micrograph shown in Fig. 7c.

In the next example, we demonstrate the applicability of the method to SFs in a more complex crystal structure, namely the sphalerite structure of GaP, which is also adopted by many other compound semiconductors. The stacking faults in this sample were formed by agglomeration of intrinsic point defects resulting from the high-concentration in-diffusion of the dopant element Zn [64]. Identifying the nature of the stacking faults—intrinsic or extrinsic—is crucial for determining the dominant diffusion mechanism, which in the present case is the so-called kick-out mechanism, whereby in-diffusing interstitial Zn atoms replace (kick out) substitutional Ga atoms, leading to a supersaturation of Ga interstitials. These interstitials agglomerate to form extrinsic stacking faults via the nucleation and growth of Frank partial dislocation loops. To maintain stoichiometry, an equal number of P atoms is incorporated into the stacking faults, with the P atoms supplied from tetrahedral voids that form in tandem with the stacking faults and contain residual Ga and partially also Zn (see [64] and Supplementary Fig. 11).

Fig. 8a shows a schematic illustration of the sphalerite unit cell consisting of two fcc sublattices, as well as its [110] projection. Analogous to Fig. 2, the projected structures of inclined ISF and ESF are shown below. (The polarity of the sphalerite structure may also be inverted, i.e., with the P columns to the right of the Ga columns instead of the left, but the translation between reference and shifted lattice will be identical.) The SF analysis can be performed by considering only one of the sublattices and ignoring the other. In this case, the Ga sublattice is highlighted as it exhibits stronger Z-contrast in HAADF-STEM micrographs.

The specimen region shown in Fig. 8b (foil thickness ≈150 nm) contains both an inclined and an edge-on SF. High-resolution imaging (Fig. 8c,d) reveals that both faults are extrinsic, as expected for a supersaturation of self-interstitials arising from the kick-out diffusion mechanism. Furthermore, the bounding dislocation of the edge-on SF is identified as a Frank partial (Fig. 8e), confirming that the fault formed via dislocation climb—i.e., through agglomeration of interstitials—rather than dislocation glide.

The inclined SF shown in Fig. 8c was further analyzed using CBED, confirming its extrinsic character. In addition, the tetrahedral features exhibiting bright contrast in Fig. 8b were analyzed by energy-dispersive X-ray spectroscopy (EDXS), revealing that some are voids partially filled with Ga, while others also contain Zn. Moreover, an inclined SF at an even higher foil thickness of ≈270 nm was successfully analyzed based on a HRSTEM micrograph. These results are presented in Supplementary Fig. 11, together with an image simulation of an inclined ESF in the sphalerite structure.

As briefly mentioned above, instead of considering the projected fault structure visible in the overlap region, it is also possible to determine the type of an inclined SF by measuring the net translation between reference and shifted lattice, e.g. by geometric phase analysis (GPA). This method is demonstrated in Supplementary Fig. 12 for the inclined SF shown in Fig. 8c as well as the one at ≈270 nm specimen thickness shown in Supplementary Fig. 11f,g.

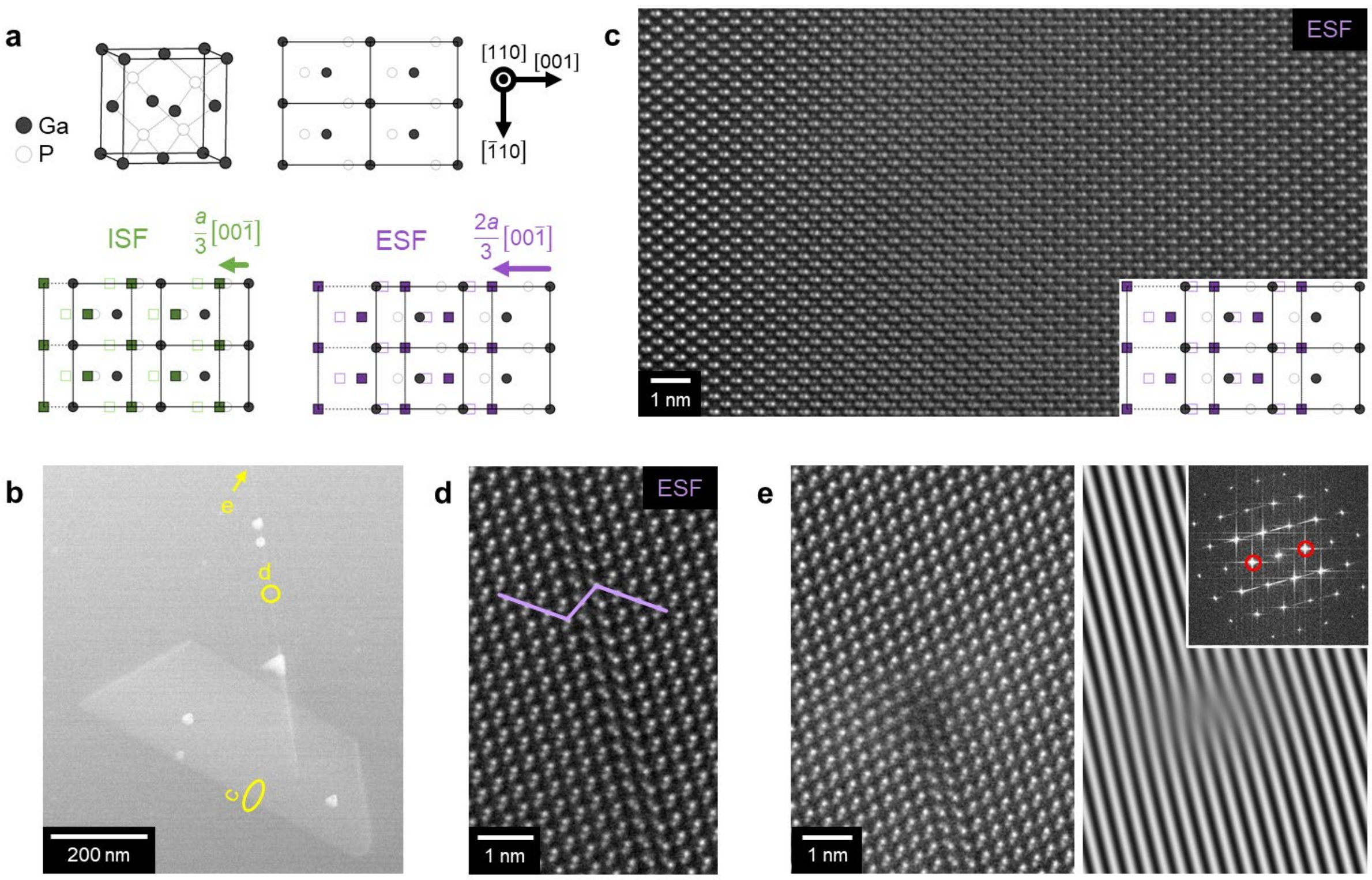

Fig. 8: Analysis of inclined and edge-on SFs in GaP with sphalerite structure. (a) Schematic drawing of the unit cell and its [110] projection, along with the expected projected structures of intrinsic and extrinsic SFs. (b) ADF-STEM micrograph of the investigated region containing both an edge-on and an inclined SF as well as tetrahedral voids which are partly filled with Ga and Zn (cf. Supplementary Fig. 11). (c) HRSTEM micrograph of the inclined SF marked in (b), revealing it to be extrinsic. (d) HRSTEM micrograph of the edge-on SF marked in (b), revealing it to be extrinsic. (e) HRSTEM micrograph of the bounding partial of the edge-on SF. As illustrated by the Bragg-filtered image using the $\pm(\bar{1}11)$ reflections marked in the Fourier transform of the micrograph (see inset), this dislocation is a Frank partial.

As briefly touched upon in the example above, to understand how a fault has formed, the partial dislocations bounding it can be analyzed. This principle can also be applied to SFs formed during mechanical testing as a result of dislocation glide. Analyzing bounding partials of edge-on faults in this way is a well-established method; as will be shown in the following, the same kind of analysis is also possible for inclined fault geometries.

Fig. 9 presents the investigation of an inclined SF formed during deformation of ERBOCo-4; specifically, in Fig. 9a, the fault's leading end, i.e., the dislocations responsible for its lengthening, is imaged in the $L1_2$ structure. The specimen region where this micrograph was recorded is shown in Fig. 9b, and the observed fault geometry is schematically drawn in Fig. 9c.

Visible from top to bottom in Fig. 9a is a region of perfect lattice, followed by a short intrinsic SF segment and finally an extrinsic fault, as evident from the projected fault structures. Leading both fault segments are the partial dislocations forming them. To determine their Burgers vectors **b** (BVs), two Burgers circuits have been drawn around them, in green (only the upper partial leading the intrinsic segment) and in purple (both partials). Due to the localized nature of atomic-column contrast formation confined to the top few nm of the foil, it is sufficient to circumnavigate a few nm of the inclined dislocation line (before it becomes invisible), avoiding the region distorted by its strain field, to obtain the Burgers circuit.

As can be seen in the magnified view in Fig. 9d, the Burgers circuit around the upper partial exhibits a closure failure equal to that partial's BV projected into the $(001)$ image plane, $\mathbf{b}_{1,\text{proj}}$, of $^{a}/_{6}\,[110]$ (cf. Supplementary Fig. 1). For the circuit around both partials, the closure failure is $^{a}/_{3}\,[110]$, representing the sum of both projected BVs $\mathbf{b}_{1,\text{proj}} + \mathbf{b}_{2,\text{proj}}$. Therefore, both projected BVs have identical values of $^{a}/_{6}\,[110]$. Given that the BV of a glissile dislocation lies in its glide plane, the true BV can be directly determined from this single projection: In inclined fault geometry, there is an unambiguous correspondence between the true BV in the glide plane and its projection in the image plane (cf. Supplementary Fig. 1); this is an advantage over edge-on analysis, where different BVs in a given glide plane may have the same projection. Here, the projected values of $^{a}/_{6}\,[110]$ correspond to BVs $\mathbf{b}_1$ and $\mathbf{b}_2$ of $^{a}/_{6}\,[11\bar{2}]$ in the $(111)$ glide plane, i.e., both dislocations are identical Shockley partials. While a conclusive determination of the BVs is possible for the glide process characterized here, it should be noted that, in the general case, an analysis based on a single projection may be ambiguous. This is especially relevant for dislocations moving by climb processes, where the BV may point out of the glide plane. A possible solution is to image the same fault configuration in two projections [47]. The external compressive stress, which has acted in [001] direction perpendicular to the image plane, results in a force on the dislocations pointing upwards in Fig. 9a, i.e, these dislocations have lengthened the SF and are therefore confirmed to be its leading configuration.

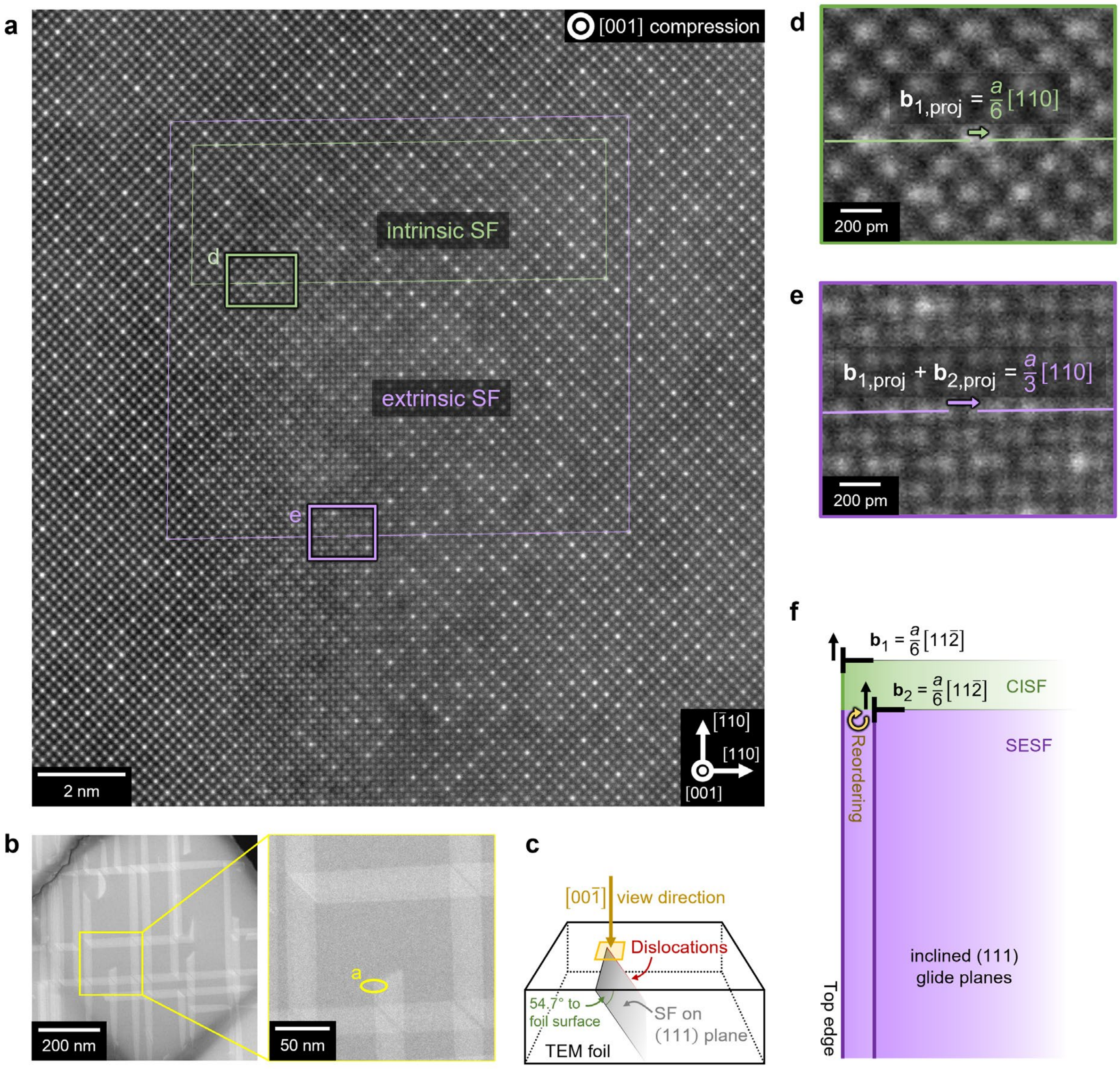

Fig. 9: Analysis of the bounding partial dislocations of an inclined SF in ERBOCo-4. (a) HRSTEM micrograph of an inclined SF containing its leading partial dislocations. Two Burgers circuits are drawn in green and purple. (b) ADF-STEM micrograph of the location where (a) was recorded (marked in yellow). (c) Schematic illustration of the observed fault geometry. (d,e) Magnified parts of (a) from the regions marked by rectangles, showing the closure failure of the Burgers circuits and thus revealing the projected Burgers vectors $\mathbf{b}_{proj}$ of the dislocations. (f) Schematic illustration of the SESF formation mechanism observed here, the so-called Kolbe mechanism (see text for details).

A schematic drawing of the observed fault configuration is shown in Fig. 9f: The first partial creates a CISF (as opposed to an SISF, whose translation vector would be different). The second partial follows on the neighboring plane, leading to the formation of the observed extrinsic fault. Overall, the observed fault structure is an instance of the well-studied Kolbe mechanism for SESF formation [19, 47, 65]. Notably, shearing by the two Shockley partials on its own creates a high-energy CESF which would not be expected to be present in an extended form. Instead, the energy of the extrinsic fault is significantly reduced by a subsequent diffusion-mediated reordering step directly behind the second partial, leading to an effective shift in superstructure by $^{a}/_{2}\langle 110\rangle$ only on the central fault plane and transforming the fault into an SESF. While not directly visible in the present analysis, this reordering mechanism has been extensively discussed [19] and later experimentally confirmed [47] after its first proposition by Kolbe [65].

This example demonstrates that the high-resolution analysis not only of SFs, but also of their bounding partials, can be performed for inclined fault geometries, making it possible to unambiguously determine their Burgers vectors from a single HRSTEM micrograph and to uncover the mechanism by which the SF has formed.

## Quasi-ultrathin foil effect

A striking feature of HRSTEM micrographs of inclined SFs is the distinct increase in contrast variation between atomic columns to the right of the fault's top edge. It is visible both as a random distribution of brighter columns in the fcc structure (e.g., Fig. 3a,b) as well as the increased visibility of the superlattice ordering in the $L1_2$ structure (e.g., Fig. 6b). The effect still appears at high foil thickness where columns in the perfect crystal exhibit significantly less distinct superlattice contrast.

This effect can be understood by the mechanism of contrast formation described in Section 2.2. If channeling is interrupted after a short distance, the difference in $Z$-contrast between atomic columns in the upper crystal is enhanced: the cross-talk between neighboring columns, which increases with increasing thickness [66] and tends to average out intensity differences between those columns, is minimized. Effectively, the HAADF contrast from the region of the foil where the SF is near the top surface resembles that of a foil only a few nanometers thick. This phenomenon, which will be referred to as quasi-ultrathin foil effect in this section, is investigated exemplarily in Fig. 10 for both the fcc and the $L1_2$ structure at a foil thickness of ≈60 nm.

In the fcc structure, the described effect makes the random distribution of high-$Z$ atoms in the solid solution clearly visible (Fig. 10a). To quantify the change in contrast, the sum intensity of each atomic column in the image was calculated. Columns were then grouped by (220) planes (see inset); within each such plane, as a measure of contrast variation, the coefficient of variation (CV), i.e., the standard deviation of the atomic column intensities for a given plane normalized by their mean intensity, was determined. These CV values for each plane are plotted below the micrograph in Fig. 10a against the depth of the SF in the foil at that plane (as calculated from the crystallographically defined angle between fault plane and ZA [34]). The area shaded in gray marks the sample region with overlapping contrast between the reference and shifted lattices, which has been excluded from the analysis. The horizontal line drawn in the plot represents the baseline CV obtained from (220) planes in the perfect crystal to the left of the SF. Where the SF is near the top of the foil, an increase in intensity variation by up to 99 % is clearly revealed.

The schematic side-view illustration above the micrograph visualizes the different sample regions present in the imaged area: perfect crystal without SF (1), region of visible overlap between reference and shifted lattice (2), quasi-ultrathin region with inclined SF at low depth, where only the part above the SF (hatched area) is visible with enhanced $Z$-contrast (3), and region with inclined SF at larger depth (4).

At a different location within the quasi-ultrathin region along the SF, still within the $\gamma$ phase, enhanced $Z$-contrast reveals a cluster exhibiting superlattice ordering. The corresponding HRSTEM image and CV analysis are shown in Supplementary Fig. 13. The image clearly identifies the cluster as a nano-precipitate of the $\gamma'$ phase, leading to alternating high and low CV values of (220) planes. This feature is likely to be a nano-scale tertiary $\gamma'$ precipitate that formed during cooling from creep temperature, as a result of oversaturation of $\gamma'$-forming elements in the $\gamma$ matrix. Without the quasi-ultrathin foil effect, this feature would likely remain undetected.

The effect of increased $Z$-contrast from the fcc solid solution was also reproduced in multislice image simulations (see Supplementary Fig. 14).

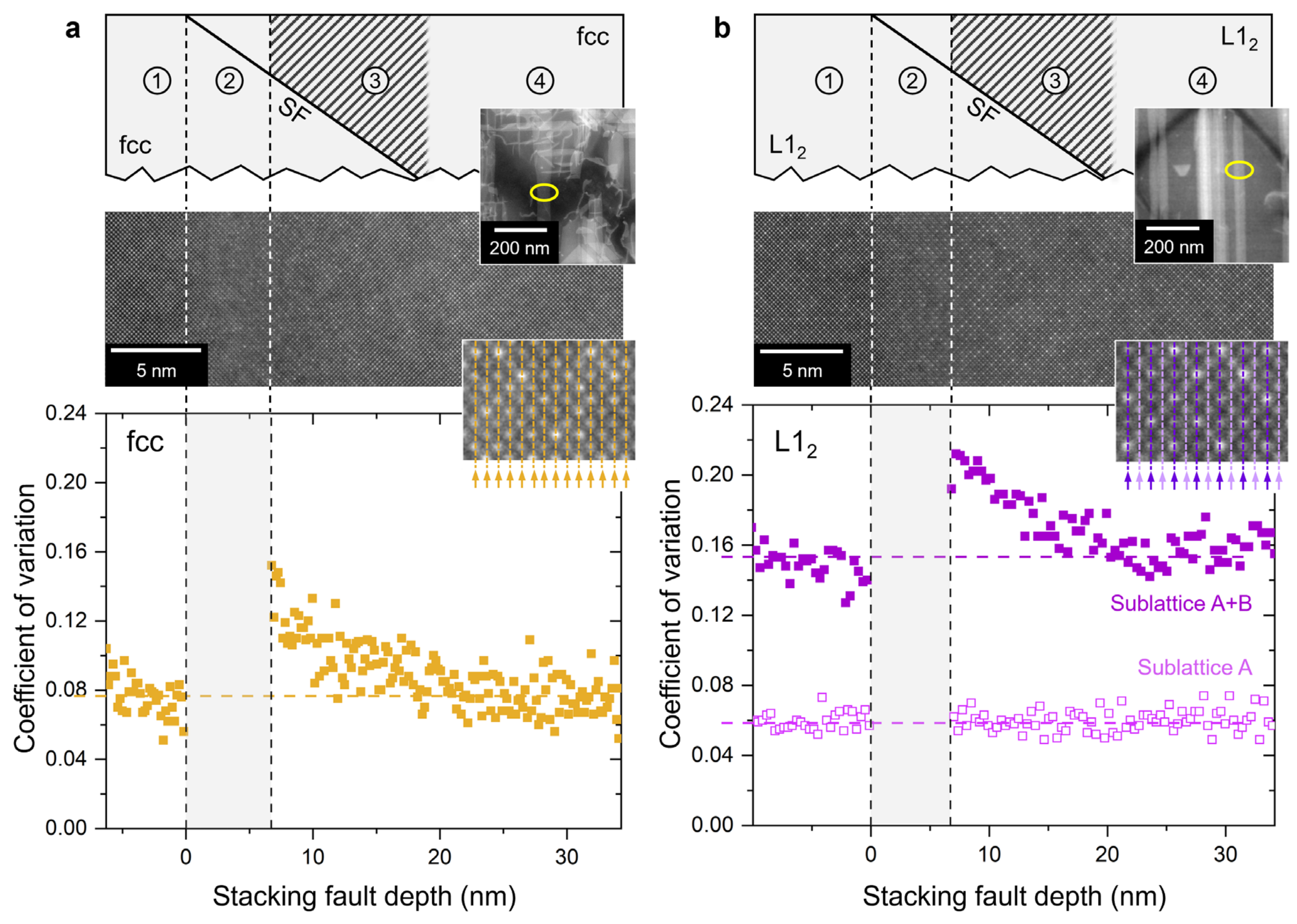

Fig. 10: Analysis of enhanced *Z*-contrast above inclined SFs. (a) HRSTEM micrograph of an inclined SF in the fcc structure. The plot below shows the CV of atomic-column intensities within each (220) plane (see inset). The horizontal line represents the baseline CV determined in a region of perfect lattice to the left of the SF. (b) Analogous evaluation of an inclined SF in the $L1_2$ structure. For this plot, (220) planes containing columns of only one sublattice and those containing both sublattices (see lower inset) were considered separately. The HRSTEM images shown in (a,b) were stitched together from two separate micrographs, respectively, to obtain a field of view containing regions to the left and right of inclined SFs, from the regions shown in the ADF-STEM micrographs (see upper insets). Above each HRSTEM micrograph, a schematic side-view illustration visualizes the region of perfect crystal (1), region of visible overlap between reference and shifted lattice (2), quasi-ultrathin region with inclined SF at low depth (3), and region with inclined SF at larger depth (4). The hatched area (3) represents the sample region contributing to the enhanced *Z*-contrast.

Contrast variations near an SF in the $L1_2$ structure have been evaluated (Fig. 10b) in the same way as for the fcc structure. Due to the superlattice ordering in this structure, two different types of (220) planes appear alternatingly, as highlighted in the inset: those containing only columns of the A sublattice (arrows and dashed lines in lighter color) and those containing alternating columns of the A and B sublattice (darker color). For the plot, both types of planes were considered separately. It is evident that contrast variations in the planes containing only the A sublattice remain negligible across all SF depths. This is related to the similar atomic numbers of Co and Ni, which occupy the A sublattice in the CoNi-base superalloy ERBOCo-4. In contrast, the CV of the mixed planes increases significantly by up to 38 % at low SF depths. This is attributed to the reduced cross-talk of the electron probe between neighboring atomic columns, which enhances the contrast between low-$Z$ (Al/Cr/Ti) and high-$Z$ (Ta/W) elements on the B sublattice. Additionally, due to the systematic variations in intensity of the two different sublattices, the baseline of the CV is significantly higher than for the A sublattice and the fcc structure.

These brief examples show that the contrast observed at inclined SFs provides opportunities for analysis even beyond characterization of the SF itself: the possibility to investigate atomic-column contrast from a region only a few nm thick without requiring an extremely thin foil may prove useful for future high-resolution studies, e.g., of long-range ordering, site occupancy, or clustering phenomena. Furthermore, in the context of the quasi-ultrathin foil effect, we suggest to also consider other two-dimensional defects such as grain or phase boundaries as suitable sites for dechanneling.

In summary, our novel approach to stacking fault (SF) analysis using high-resolution scanning transmission electron microscopy (HRSTEM) of inclined faults overcomes the long-standing limitations of the two classical SF analysis methods that have been employed for decades. Directly applicable as a complementary tool to edge-on HRSTEM fault analysis, the new method is highly flexible and robust in determining the intrinsic or extrinsic character of stacking faults, as demonstrated in both the fcc and $L1_2$ crystal structures of a compression-deformed superalloy, as well as an oxide-dispersion strengthened medium-entropy alloy and in a semiconductor specimen with sphalerite structure. This method is applicable to both [110] and [001] zone axis projections and relies on evaluating the relative displacement between the projected crystal lattices above and below the stacking fault. These lattices appear directly superimposed in HAADF-HRSTEM micrographs near the intersection of the stacking fault with the upper surface of the TEM lamella. We showed that the method remains applicable even at foil thicknesses typical for conventional TEM (>100 nm) and in cases where SFs significantly overlap in projection, as only the contrast formed near the top edge of the SF is relevant for the analysis. Furthermore, the investigation of partial dislocations bounding the SF was demonstrated to be possible even for inclined fault geometries. Multislice probe-propagation simulations provided deeper insight into how the SF-induced shifts of atomic columns within the foil influence contrast formation in HRSTEM lattice images. Most importantly, when combined with the well-established HRSTEM analysis of edge-on SFs, the proposed method enables, for the first time, reliable analysis of SFs on all four {111} glide planes in a single sample using one microscopy technique. Although the fcc and related structures are particularly suitable for this method owing to the presence of inclined glide planes in the projections commonly used for HRSTEM analysis, the underlying principle is transferable to other crystal structures as well and is expected to facilitate defect characterization across a wide range of material systems, including high- and medium entropy alloys, steels, and thin-film systems.

Furthermore, we have shown that the HAADF contrast from the region where the inclined SF approaches the top of the foil closely resembles that of an ultrathin lamella, enabling studies which would traditionally require a very low foil thickness of only a few nanometers. We anticipate that this quasi-ultrathin foil effect will have significant practical relevance, for example in studies of long-range ordering, compositional fluctuations, and nanoclustering. We propose extending this concept to other two-dimensional defects, such as grain boundaries or phase boundaries, where de-channeling effects are expected to be even more pronounced than in the case of SFs.

## Methods

### Specimen preparation

For the experimental investigation of the inclined SFs, the single-crystalline CoNi-base superalloy ERBOCo-4 (composition in at.%: $Co_{43.2}Ni_{32}Al_8W_{5.7}Ti_{2.8}Ta_{1.8}Cr_6Si_{0.4}Hf_{0.1}$) [67] was cast using the Bridgman process. To obtain the γ/γ′ microstructure, a three-step heat treatment (8 h at 1280 °C, 5 h at 1050 °C, 16 h at 900 °C) was performed. Specimens for compression testing with a diameter of 3 mm and a height of 4.5 mm were fabricated by wire electrical discharge machining. Constant strain-rate compression testing at strain rates of $10^{-3}$ and $10^{-5}$ $s^{-1}$ along the [001] direction was performed at an Instron 4505 electromechanical universal testing machine at a temperature of 850 °C up to a plastic strain of 1 %. [001] and [110] sections were cut from the specimens and ground down to a thickness of 100-150 µm. To attain electron transparency, samples were electrolytically thinned in a solution of 16.7 % nitric acid and 83.3 % methanol in a Struers Double Jet Tenupol-5 preparation system at 40 to 45 V and a temperature of −25 °C. Additional details about the alloy fabrication, deformation, and TEM sample preparation can be found in an earlier publication [27].

If needed, electropolished specimens were subjected to an Ar-ion polishing step to further reduce foil thickness. This step was performed using a Gatan PIPS II first at 3 kV, then at 1 kV, at an incidence angle of 6 °.

The focused ion-beam lift-out and thinning of the lamella for the surface-quality comparison was performed at an FEI Helios Nanolab 660 instrument using a $Ga^{+}$ ion beam.

GRX-810 (composition in wt.%: Ni-33Co-29Cr-3W-1.5Re-0.75Nb-0.3Al-0.25Ti-0.05C) powder was coated with 1 wt.% $Y_2O_3$ by resonant acoustic mixing (RAM), followed by laser powder bed fusion (LPBF) processing and hot isostatic pressing (HIP). Detailed RAM, LPBF, and HIP parameters are reported in [60, 62, 63]. Sections perpendicular to the build direction were mechanically ground to 100–150 µm and subsequently electropolished (TenuPol-5) in a 5 vol.% perchloric acid − 95 vol.% methanol mixture at 25 V and -25 °C.

The GaP sample studied in Fig. 8 and in Supplementary Fig. 11 were taken from a sample series in which the dopant element Zn was indiffused into GaP(001) single crystals using argon-flushed, evacuated and sealed quartz ampoules with elemental Zn as diffusion source [68]. In specific regions of the diffusion profile, interstitial-type dislocation loops formed as a result of supersaturation of Ga self-interstitials generated by the kick-out diffusion of Zn. Among these loops are faulted loops on {111} planes, as investigated in Fig. 8 and in Supplementary Fig. 11, which are bounded by Frank partial edge dislocations. The HRSTEM and CBED analyses were carried out on [110] cross section samples prepared by grinding, polishing and Ar ion beam milling. Further details on these samples and their preparation can be found in Ref. [64] and references therein.

## Transmission electron microscopy

HRSTEM imaging was performed using a probe-corrected Thermo Fisher Scientific Spectra 200 C-FEG at 200 kV using the high-angle annular dark field (HAADF) detector and a semi-convergence angle of 30 mrad (unless noted otherwise). For the camera length, values between 62 mm and 198 mm were used, corresponding to collection angle ranges ranging from 90-200 mrad to 28-169 mrad, i.e., HAADF or annular dark-field (ADF) imaging conditions. EDXS measurements using the Super-X G2 detector were performed at the same microscope. EDXS data was evaluated using the Thermo Fisher Scientific Velox software. HRSTEM investigations of the ODS-containing GRX-810 alloy were conducted using a probe-corrected Thermo Fisher Scientific ThemisZ operated at 200 kV. HAADF imaging was performed using a semi-convergence angle of 30 mrad and a camera length of 91 mm. Where necessary, the influence of scan noise and drift distortions was minimized by recording multiple frames and averaging them. Furthermore, dark-field TEM imaging in a two-beam condition was performed using a Philips CM 30 TEM at 300 kV. Zero-loss filtered CBED patterns (semi-convergence angle ≈5.5 mrad) were acquired using a Thermo Fisher Scientific Titan Themis[3] TEM coupled with a Gatan Quantum ERS image filter/spectrometer. The microscope is additionally equipped with a monochromator (not excited), Cs-correctors on both image and probe-forming side, and operated at 300 kV. The flatness of the energy loss plane (i.e., isochromaticity) in diffraction mode was carefully aligned manually and the energy selection window was set to 10 eV around the zero-loss peak.

## Crystal files and multislice simulations

The CrystalMaker X software (version 10.8.2; CrystalMaker Software Ltd, Oxford) was utilized to generate crystal structures for illustration and image simulation purposes. The multislice simulations for the fcc structure were performed using a pure Ni crystal as a typical representative of this crystal structure, similar in atomic number to the experimentally investigated CoNi-based solid solution. For $L1_2$, the model $Co_3(Al,W)$ structure was used with a randomized occupation of the B sublattice. This represents the experimentally observed ordered structure well, which also exhibits a higher average atomic number of the B sublattice (i.e., the corners of the unit cell cube). It should be noted that there are also phases in which the A sublattice shows a higher average atomic number, such as the $Ni_3Al$ phase. In such cases, the pattern of brighter and darker atomic columns in HAADF contrast will be inverted.

Multislice simulations of STEM images as well as probe propagation were carried out using the Dr.Probe software [69] at an accelerating voltage of 200 kV, a convergence semi-angle of 30 mrad (unless noted otherwise), and with no aberrations. HAADF image simulations were performed with an angular detection range of 90-200 mrad, a pixel size of 10 pm, and an effective source size of 0.02-0.025 nm.

### Determination of intensity variation

Code for the evaluation of atomic-column contrast was written in Python as a Jupyter Notebook [70] using the Hyperspy and Atomap packages [71, 72]. To quantify the increased contrast close to the intersection between the inclined SF and the top of the foil, the integrated intensity of each atomic column was calculated and its position within the micrograph was determined. Subsequently, atomic columns were grouped according to their distance perpendicular to the intersection of the inclined SF and the top of the foil, i.e., along (220) planes (the vertical planes highlighted by arrows and dashed lines in the inset of Fig. 8). The mean $\mu$ and standard deviation $\sigma$ of each group were calculated and the coefficient of variation (CV) was determined by CV = $\sigma/\mu$. Code generation was assisted by the ChatGPT software (models 4o and o3-mini-high; OpenAI, San Francisco, United States). The analyzed micrographs were manually stitched using CorelDRAW Graphics Suite 2021 (Alludo, Ottawa, Canada) to cover a sufficiently large area such that the increased CVs returned to baseline levels to the left and right of the inclined SF.

## Data availability

The electron microscope and simulation data used in this study are available at the open-access repository Zenodo under accession code 14906582 [https://doi.org/10.5281/zenodo.14906582].

## Code availability

The code used in this study is available at the open-access repository Zenodo under accession code 14906582 [https://doi.org/10.5281/zenodo.14906582].

## Acknowledgements

The authors thank Mingjian Wu for help with CBED experiments, as well as Timothy M. Smith and the National Aeronautics and Space Administration (NASA) for providing specimens of the alloy GRX-810.

## Funding statement

The authors gratefully acknowledge funding from the German Research Foundation (DFG) through projects A7 (N.K. and E.S.) and B3 (A.B. and S.N.) of the collaborative research project SFB/Transregio 103 “From Atoms to Turbine Blades - a Scientific Approach for Developing the Next Generation of Single Crystal Superalloys”. A.B. acknowledges financial support from the Alexander-von-Humboldt Foundation. A.B. and M.J.M. acknowledge the support of the NSF and the DMREF program under grant No. 2323717.

## Author contributions statement

E.S. conceived the research. E.S. and N.K. developed the methodology. N.K. performed TEM experiments and analyzed the data, with contributions from L.M. L.M. and N.K. performed multislice simulations. A.B. performed TEM experiments on the GRX-810 alloy and the analysis of intensity variations. A.B. and S.N. provided specimens of the investigated alloy ERBOCo-4. N.K. wrote the initial version of the manuscript. E.S. revised the manuscript. N.K., L.M., A.B., M.J.M., S.N., and E.S. contributed to the final version of the manuscript.

## Competing Interests Statement

The authors declare no competing interests.

## Tables

(not applicable)

# Supplementary Information

## Stacking faults from a different angle – Overcoming the edge-on limit in high-resolution defect analysis

**Nicolas Karpstein [a], Lukas Müller [a], Andreas Bezold [b,c], Michael J. Mills [c], Steffen Neumeier [b], Erdmann Spiecker [a,*]**

[a] Friedrich-Alexander-Universität Erlangen-Nürnberg, Department of Materials Science & Engineering, Institute of Micro- and Nanostructure Research, and Center for Nanoanalysis and Electron Microscopy (CENEM), IZNF, Erlangen, Germany

[b] Friedrich-Alexander-Universität Erlangen-Nürnberg, Department of Materials Science & Engineering, Institute I: General Materials Properties, Erlangen, Germany

[c] The Ohio State University, Department of Materials Science & Engineering, Columbus, OH, USA

* Corresponding author: erdmann.spiecker@fau.de

Supplementary Note 1. **Stacking fault types and corresponding lattice translations in the fcc and $L1_2$ crystal structures.**

The fcc and $L1_2$ crystal structures give rise to various SF types, which will be outlined in the following (Supplementary Fig. 1). To visualize the lattice translation associated with each type of stacking fault exemplarily in the $(111)$ glide plane, a stack of three $(111)$ planes is shown where the bottom, middle, and top layer are represented by small, medium-sized and large atoms, respectively. Arrows represent the translation of the top layer against the lower ones required to form a given fault.

In the fcc structure (Supplementary Fig. 1a), $a/_6\langle 211\rangle$ Shockley partial dislocations introduce intrinsic SFs (ISFs) into the crystal structure (green arrows) and make it appear as though a plane is missing from the …ABCABC… stacking sequence of close-packed planes. As evident from the illustration, the corresponding translation can occur in three different directions; however, all three translation vectors result in crystallographically identical fault structures. In other words, the difference vector between any two of these vectors is equal to a lattice translation vector. An extrinsic SF (ESF) consists of two ISFs on neighboring glide planes, resulting in net translations of type $a/_3\langle 211\rangle$ as illustrated by purple arrows. In this case, the stacking sequence appears to have an extra plane inserted into it. For simplicity, the arrows representing two-layer faults in Supplementary Fig. 1 are drawn in one crystal plane to visualize the total shifts between the upper and lower half crystal.

In the $L1_2$ structure (Supplementary Fig. 1d), the superstructure gives rise to additional distinctions between fault types based on whether nearest-neighbor violations (i.e., Al sites being unfavorably close to each other) exist between the fault planes. Intrinsic and extrinsic SFs still have the same effect on the stacking sequence as in the fcc structure. However, a Shockley partial now introduces a complex intrinsic SF (CISF, light green arrows) with additional nearest-neighbor violations. For a superlattice intrinsic SF (SISF, dark green arrows) without such violations, a twice as large $a/_3\langle 211\rangle$ Kear partial is required. Again, in both cases, this translation is possible in three different directions, and in the case of the SISF, all three translation directions lead to identical fault structures. On the other hand, the three CISF are crystallographically distinct, albeit only in their superstructure. Analogous to the fcc structure, a complex extrinsic SF (CESF, light purple arrows, also referred to as CESF-2) can be described as comprising two identical CISFs on neighboring planes. Similarly, a superlattice extrinsic SF (SESF, dark purple arrows) consists of two neighboring SISFs. Notably, the net translation vectors of CESF and SESF are identical; the two faults differ only in the translation of their central plane. Consequently, a CESF can be transformed into an SESF by local reordering of the central fault plane, as postulated by the Kolbe mechanism (see main text for details).

**Supplementary Note 2. Projections of fault translation vectors along the [110] and [001] viewing directions.**

For the projections of the net translation vectors of faults along the [110] and [001] viewing directions, the faults are considered to be situated on the (111) plane in accordance with Supplementary Fig. 1a,d and the main text, without loss of generality. Likewise, the structures in Supplementary Fig. 1b,c,e,f are rotated so that the depth of the SF and its (111) habit plane increases from left to right. In [110] projection, this means that the [001] direction points to the right, whereas in [001] projection, the [110] direction points to the right.

Projected net translation vectors of ISF and ESF in the fcc structure are shown in Supplementary Fig. 1b,c.

In the [110] ZA (Supplementary Fig. 1b), the projections of the ISF translation vectors $^a/_6\,[11\bar{2}]$, $^a/_6\,[1\bar{2}1]$, and $^a/_6\,[\bar{2}11]$ are $^a/_3\,[\mathit{00\bar{1}}]$, $^a/_{12}\,[\mathit{3\bar{3}2}]$, and $^a/_{12}\,[\mathit{\bar{3}32}]$, respectively (for better distinction, projected vectors are written in italics). Since all three ISF translation vectors result in the same fault structure, the same holds true for the projected vectors (summarized as $^a/_3\,[\mathit{00\bar{1}}]$ in the main text) and resulting structure, where the shifted lattice appears translated by one third of the periodicity to the left of the reference lattice (Fig. 2a, left).

The ESF translation vectors and thus their respective projections are twice as large. Again, the resulting three projected fault structures are identical (and summarized as $^{2a}/_3\,[\mathit{00\bar{1}}]$ in the main text). Here, the shifted lattice appears to the right of the reference lattice (Fig. 2a, left).

In the [001] ZA (Supplementary Fig. 1c), the ISF translation vectors $^a/_6\,[11\bar{2}]$, $^a/_6\,[1\bar{2}1]$, and $^a/_6\,[\bar{2}11]$ have projections of $^a/_6\,[\mathit{110}]$, $^a/_6\,[\mathit{1\bar{2}0}]$, and $^a/_6\,[\mathit{\bar{2}10}]$, respectively, with identical projected fault structures (the projected translation is summarized as $^a/_6\,[\mathit{110}]$ in the main text). This means that the shifted lattice appears to the right of the reference lattice (Fig. 2a, right).

Again, the three translation vectors of ESFs and their projections are twice as large and the three projected structures identical (summarized as $^{a}/_{3}\,[110]$ in the main text) with the shifted lattice appearing to the left of the reference lattice (Fig. 2a, right).

Projections of SISF and SESF translation vectors in the $L1_2$ structure are illustrated in the same way in Supplementary Fig. 1e,f. Compared to the ISF in the fcc structure, translation vectors of the SISF – and thus their projections in the [110] and [001] ZAs – are twice as large and antiparallel, also leading to projected fault structures where the shifted lattice appears to the left of the reference lattice in the [110] ZA (Fig. 2b, left) and to the right in the [001] ZA (Fig. 2b, right). The net translation vectors of SESFs in $L1_2$ are identical to those of ESFs in fcc, and the same applies to their projections. In the projected fault structures, the shifted lattice appears to the right of the reference lattice in the [110] ZA (Fig. 2b, left) and to the left in the [001] ZA (Fig. 2b, right).

**Supplementary Note 3. Projected structures of complex faults in the $L1_2$ structure.**

The difference between complex faults and their non-complex (superlattice) counterparts in the $L1_2$ structure is only in the superlattice ordering, i.e., the stacking sequence of close-packed planes, ignoring the individual sublattices, is the same. Therefore, atomic columns also appear in the same positions in HRSTEM micrographs; differences in projected fault structures can only arise in terms of which atomic columns appear brighter or darker due to Z contrast.

The projected fault translation vectors of CISFs, along with SISFs for comparison, are presented in Supplementary Fig. 2 (cf. Supplementary Fig. 1b for the in-plane view of fault translation vectors) along with HAADF-STEM multislice simulations of the corresponding faults. Unlike their superlattice counterparts, the three possible translation directions result in crystallographically distinct fault structures, again differing only in their superstructure. Thus, all three are considered separately and numbered for reference.

In the [110] ZA (Supplementary Fig. 2a), the projected fault translation vector of $^{a}/_{3}\,[00\bar{1}]$ results in a projected fault structure identical to that of an SISF (cf. Fig. 6 in the main text), where neighboring columns are of the same type and thus appear at the same brightness in HAADF contrast (Supplementary Fig. 2c). The projected translation vectors $^{a}/_{12}\,[3\bar{3}2]$ and

$^{a}/_{12}\,[\bar{3}32]$, on the other hand, result in a different projected structure where mixed neighboring columns appear. Thus, only in these two cases, the CISF can unambiguously be identified as such. As a visual guide, brighter columns are highlighted by yellow circles.

A similar pattern emerges in the [001] ZA (Supplementary Fig. 2b,d): the projected structure associated with $^{a}/_{6}\,[110]$ is identical to that of the SISF, recognizable by the presence of $(1\bar{1}0)$ planes (i.e., horizontal rows) containing only one sublattice. The other two projected translation vectors, $^{a}/_{6}\,[1\bar{2}0]$ and $^{a}/_{6}\,[\bar{2}10]$, result in projected structures characteristic to the CISF.

CESFs are not distinguishable at all from SESFs using the method presented in this work. This is because the contrast of projected fault structures almost only depends on the net translation vector of the imaged inclined fault, which is identical for CESF and SESF (cf. Supplementary Fig. 1d). Between the two fault structures, the superstructure only of the central fault plane is shifted. This difference within a single crystal plane has a negligible effect on image contrast.

## Supplementary Figures

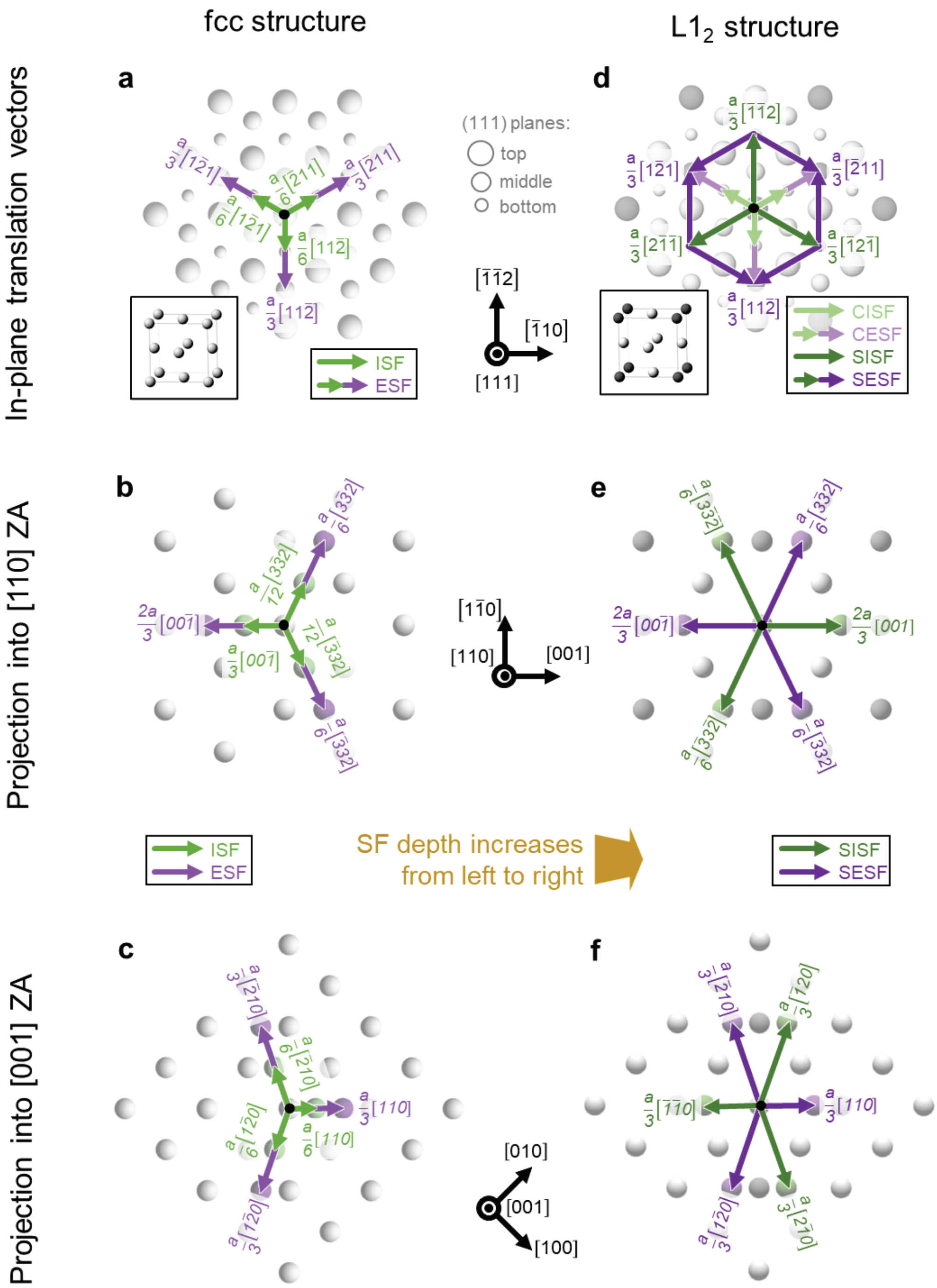

Supplementary Fig. 1: SF types, their translation vectors and their [110] and [001] projections, in the (a-c) fcc and (d-f) $L1_2$ structure. (a,d) Schematic illustrations of a stack of three (111) planes in the (a) fcc and (d) $L1_2$ structure. Unit cells are shown as an inset. The vectors by which the top half-crystal is translated to form a given fault are represented by arrows colored depending on the fault type. Schematic illustrations of the fcc and $L1_2$ crystal structures in (b,e) [110] and (c,f) [001] projection, with net translation vectors of (S)ISF and (S)ESF projected into the respective plane. The projected vectors correspond to those shown in (a,d), i.e., to a fault on the (111) plane. In accordance with the analysis presented in the main text, the structures are rotated so that the depth of the (111) plane inclined to the viewing direction increases from left to right.

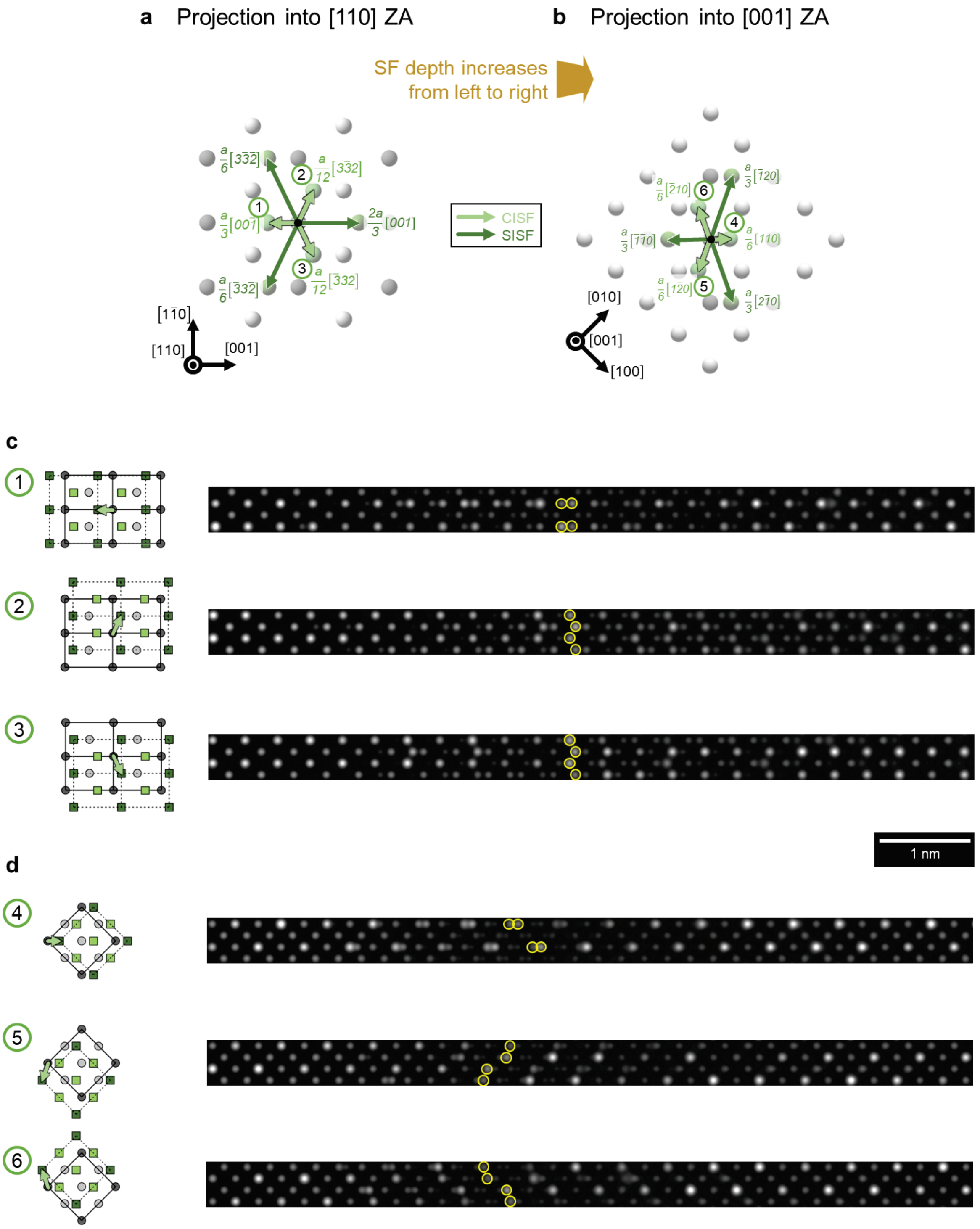

Supplementary Fig. 2: Schematic illustration of the $L1_2$ crystal structure in (a) [110] and (b) [001] projection, with net translation vectors of CISF (and SISF for comparison) projected into the respective plane. The projected vectors correspond to those shown in Supplementary Fig. 1d, i.e., to a fault on the (111) plane. In accordance with the analysis presented in the main text, the structures are rotated so that the depth of the (111) plane inclined to the viewing direction increases from left to right. Below, projected structures and multislice image simulations (for a reduced sample thickness of 12 nm for enhanced superlattice contrast) corresponding to the different CISF translation vectors are shown for (c) [110] and (d) [001] projection.

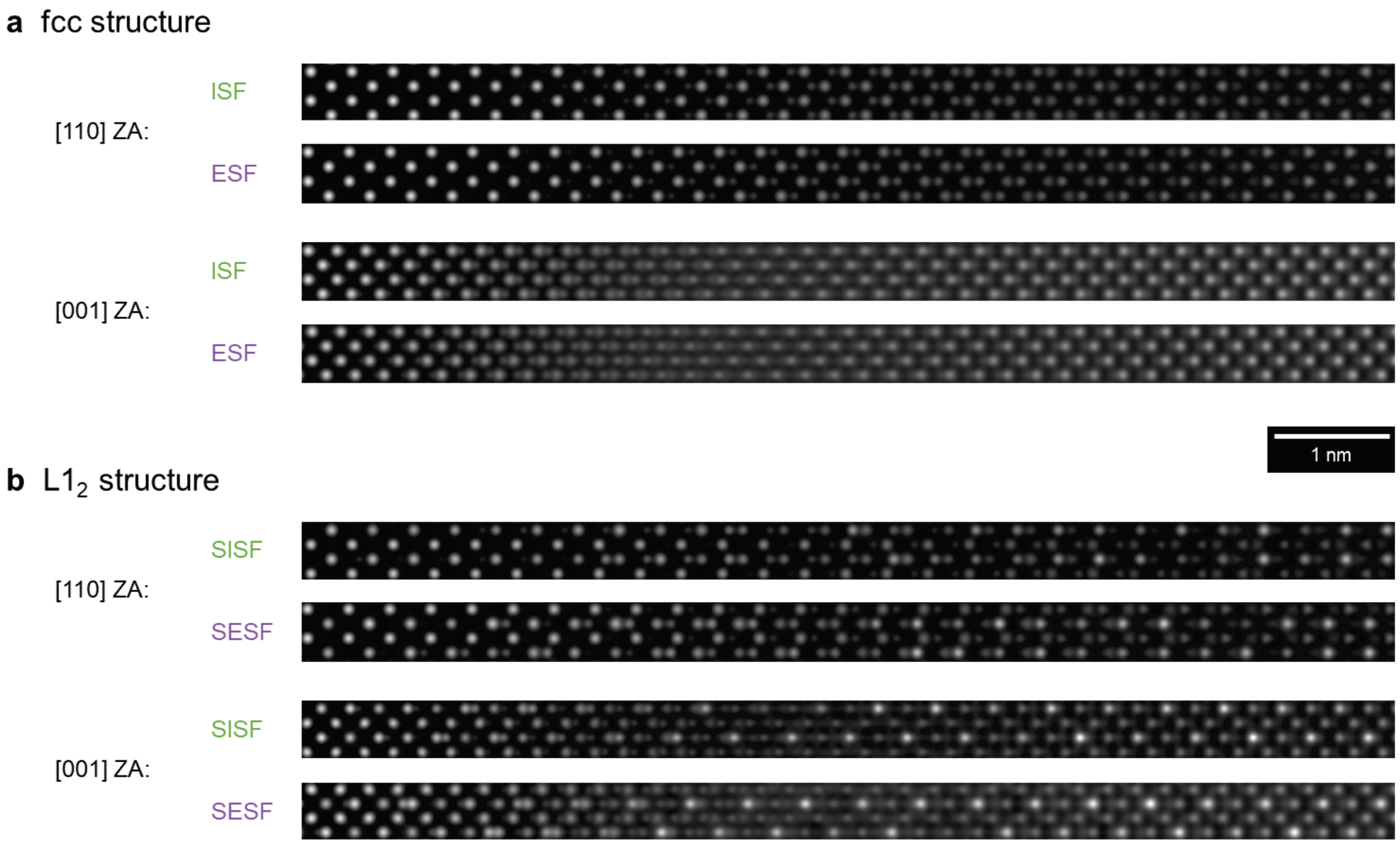

Supplementary Fig. 3: HAADF-STEM multislice simulations of inclined SFs for the different fault types in (a) fcc structure and (b) L1$_2$ structure. Foil thickness ≈30 nm in all cases.

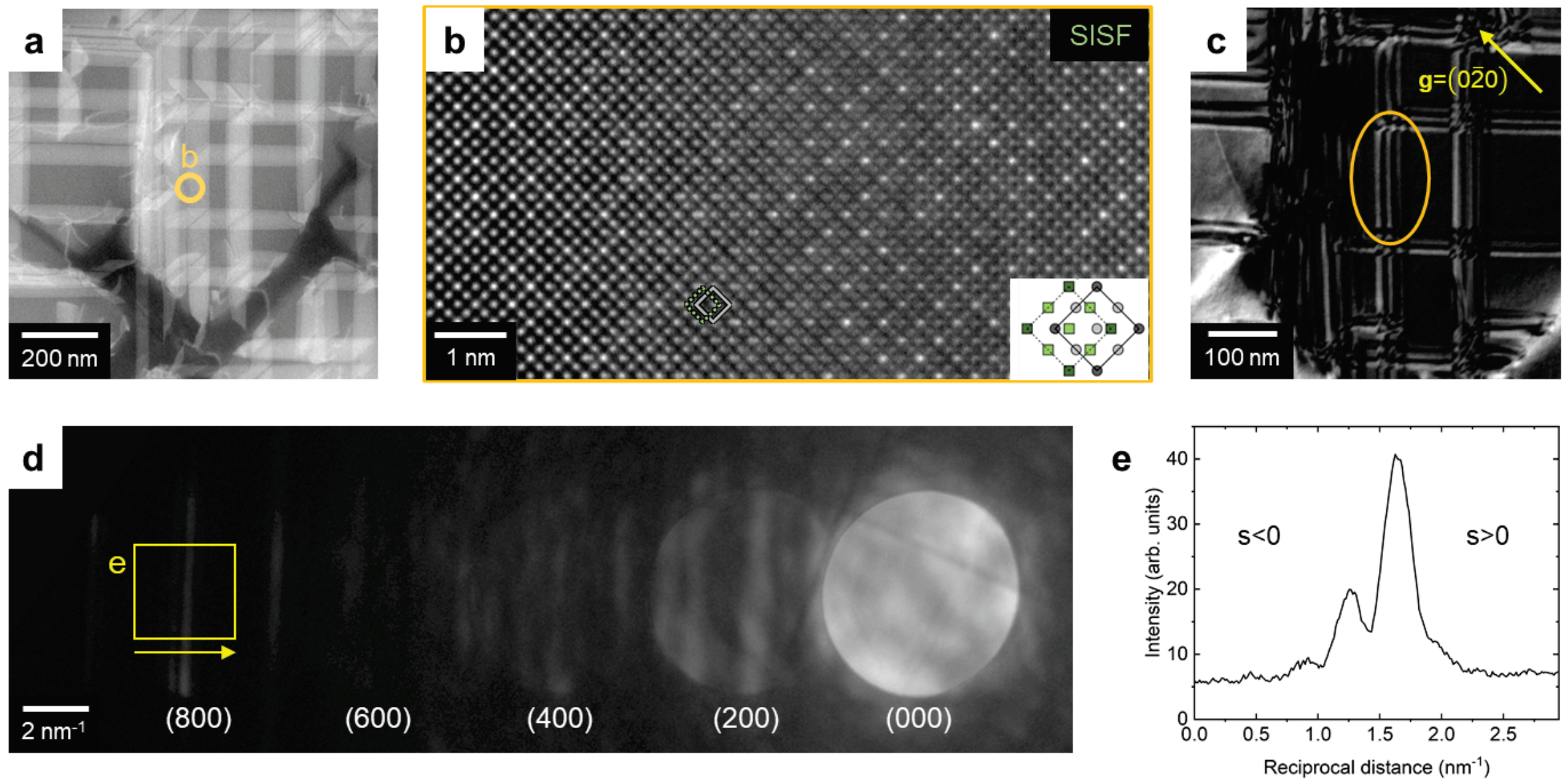

Supplementary Fig. 4: Analysis of the same SF using three different methods. (a) ADF-STEM micrograph of the surrounding microstructure. (b) HRSTEM micrograph of the top edge of the fault marked in (a), revealing it to be intrinsic. (c) Dark-field TEM micrograph of the same fault in a (0$\bar{2}$0) two-beam condition. **g** points towards the bright outer fringe, confirming the fault's intrinsic character. (d) Energy-filtered CBED pattern with the beam converged on the center of the same fault and the (800) reflection excited. (e) Intensity profile of the region marked in (d), clearly revealing splitting of the (800) Bragg line. The additional line of lower intensity appears in the direction of negative excitation error, indicating a phase shift of $-{}^{2\pi}/_{3}$ and thus an intrinsic fault.

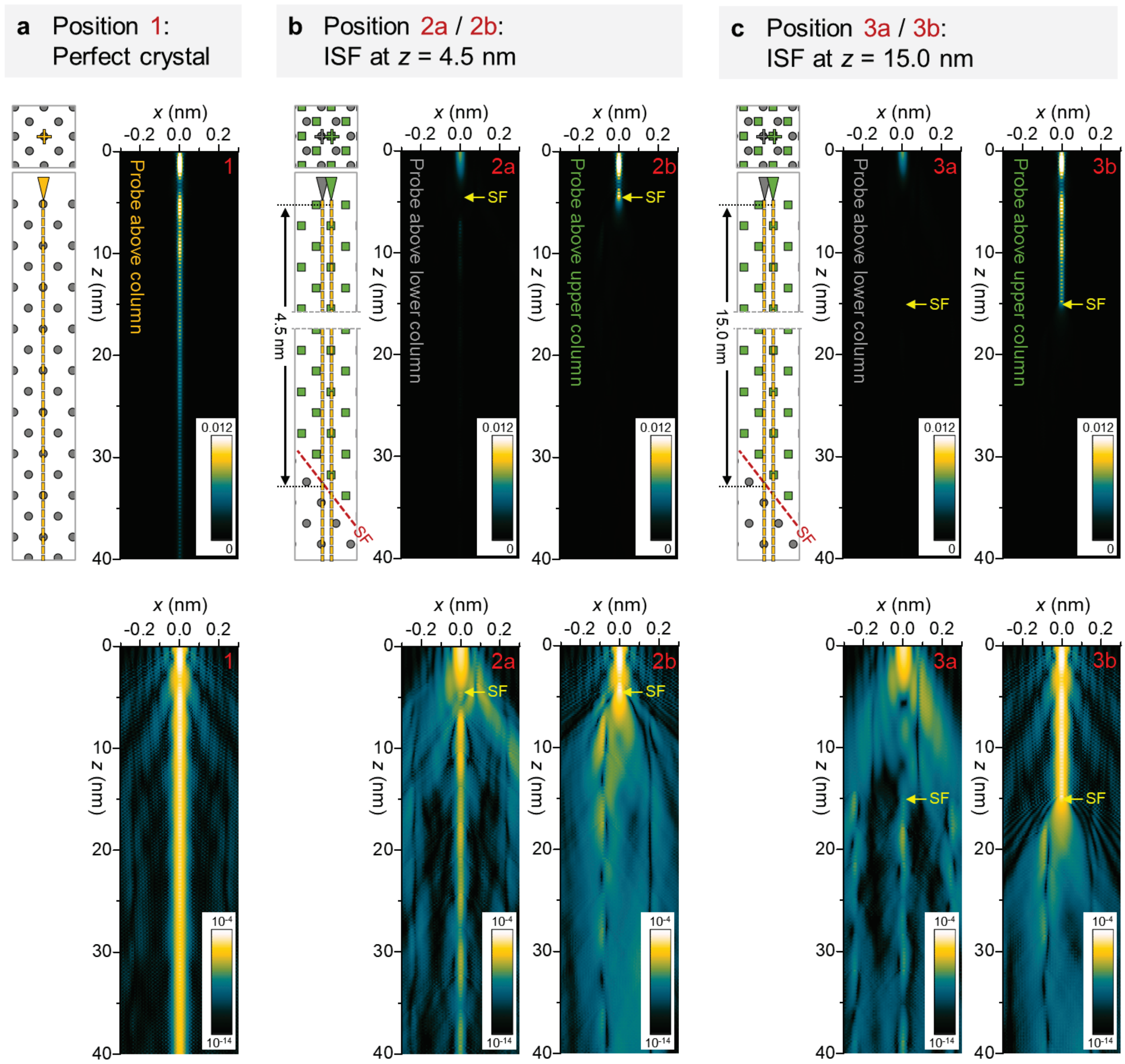

Supplementary Fig. 5: Linear (top) and logarithmic (bottom) presentation of the slices of simulated probe intensity distributions shown in Fig. 4 in the main text.

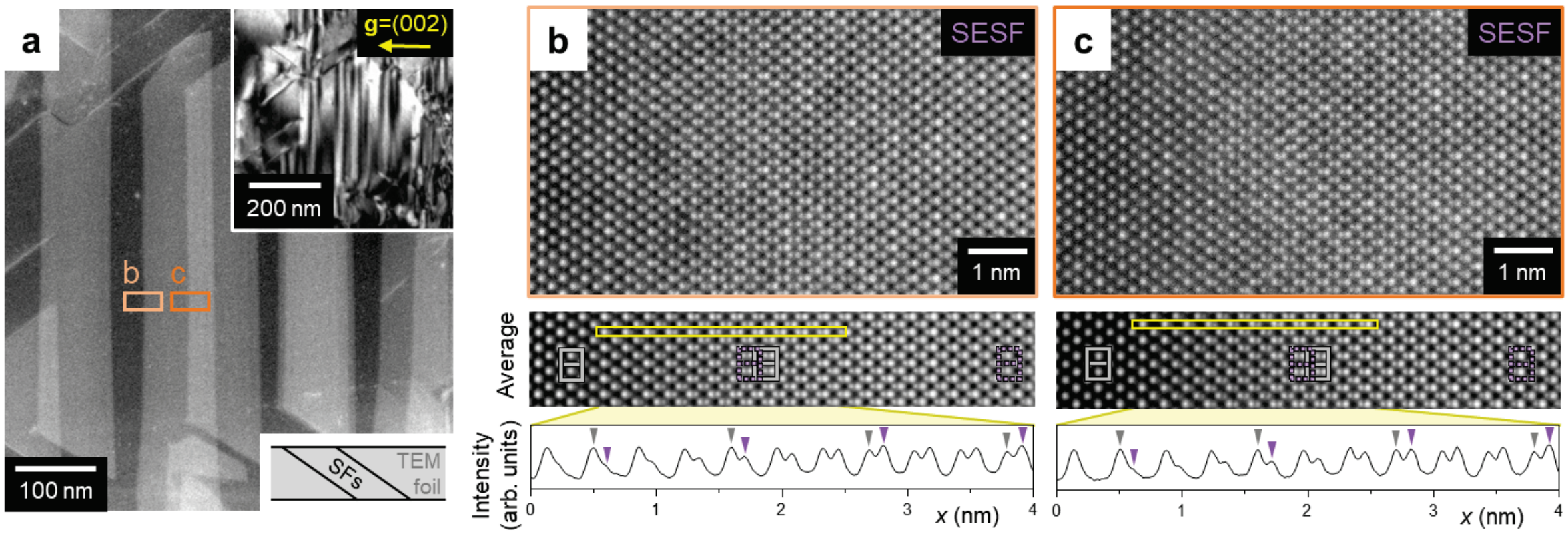

Supplementary Fig. 6: Analysis of SFs overlapping in projection. (a) ADF-STEM micrograph of overlapping SFs in the [110] ZA. The inset at the top right shows a dark-field TEM image of the same region with fringe contrast difficult to interpret. (b,c) HRSTEM micrographs of the locations marked by rectangles in (a), from which both faults can be analyzed separately and identified as extrinsic. Below each HRSTEM micrograph, the periodically averaged signal is shown along with intensity linescans from the areas marked by yellow rectangles. In the intensity profiles, some peaks of the reference and shifted lattice are highlighted by grey and colored triangles, respectively.

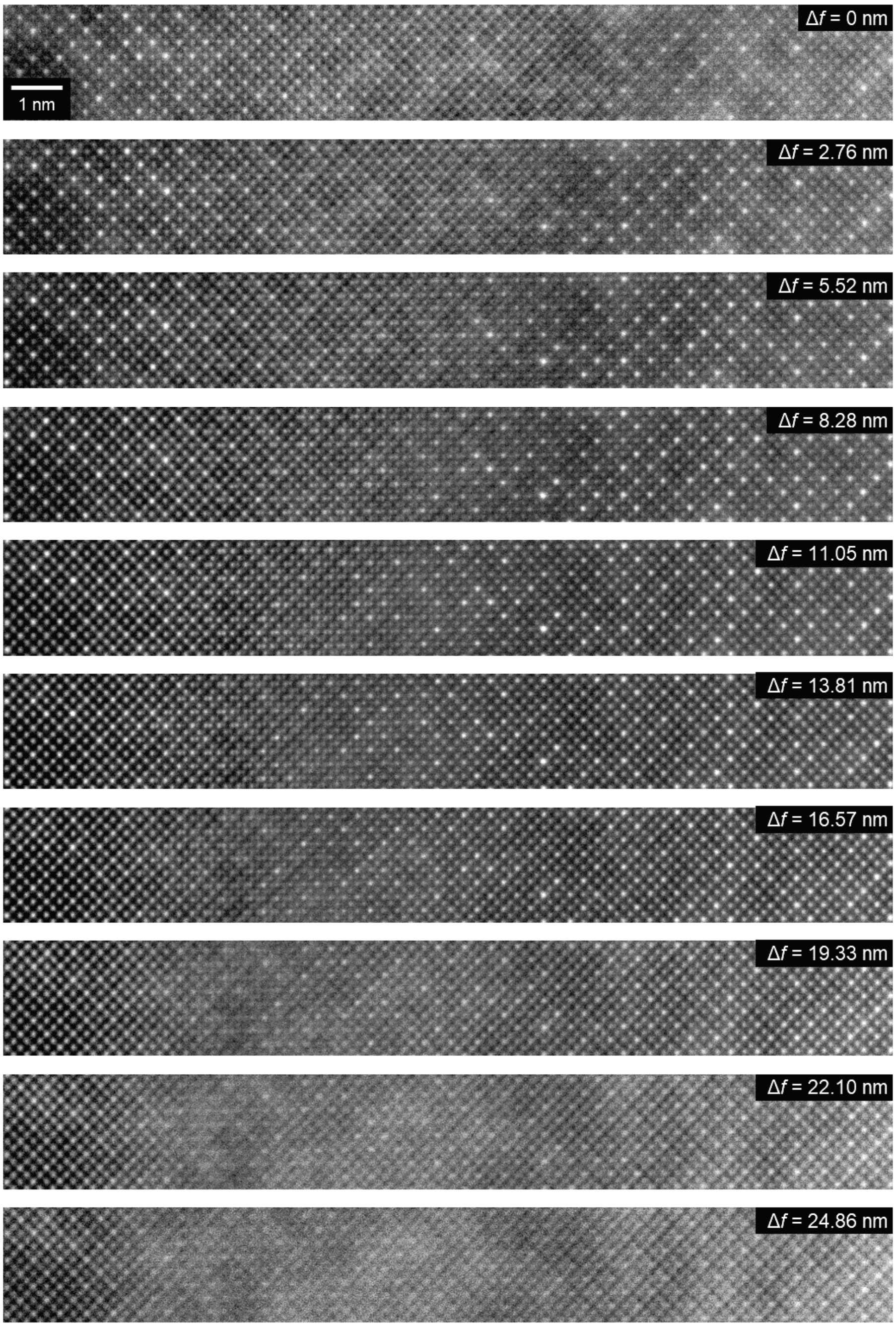

Supplementary Fig. 7: HRSTEM micrographs of the same specimen region containing the top edge of an inclined SISF, under varying defocus $\Delta f$ (arbitrary reference value) and at a constant beam semi-convergence angle $\alpha$ of 30 mrad. With increasing overfocus (higher $\Delta f$), the thickness range sampled by atomic-column contrast is shifted more towards the top of the foil, making the region of overlap between reference and shifted lattice appear further to the left (and vice versa for increasing underfocus).

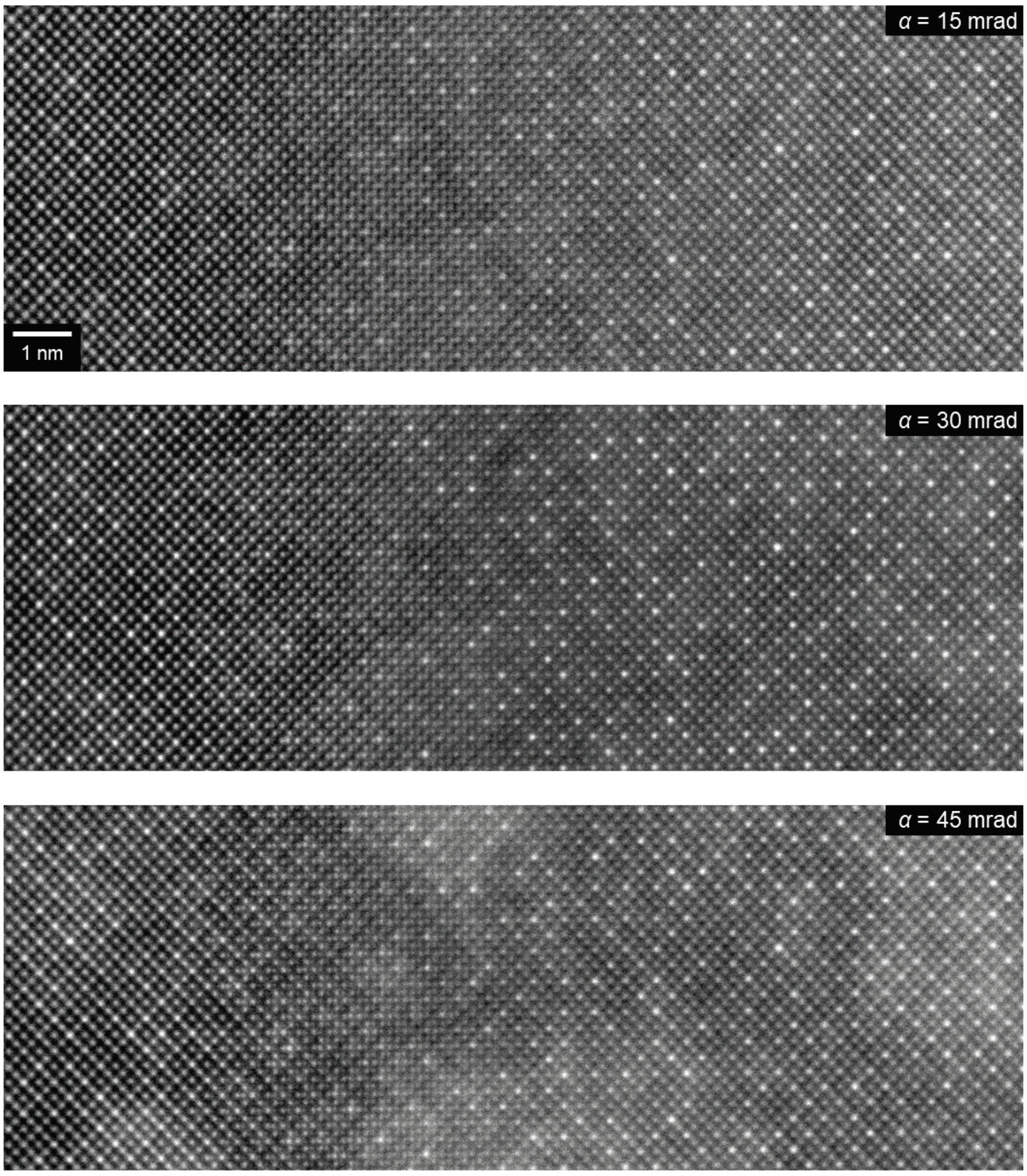

Supplementary Fig. 8: HRSTEM micrographs of the same specimen region containing the top edge of an inclined SISF, under varying beam semi-convergence angle $\alpha$ (achieved by changing the C2 aperture). The higher $\alpha$, the lower the depth of field; thus, the region of visible overlap between reference and shifted lattice becomes narrower. Note that no re-tuning of the probe corrector was performed after changing $\alpha$, leading to an increased effect of aberrations in the lower micrograph.

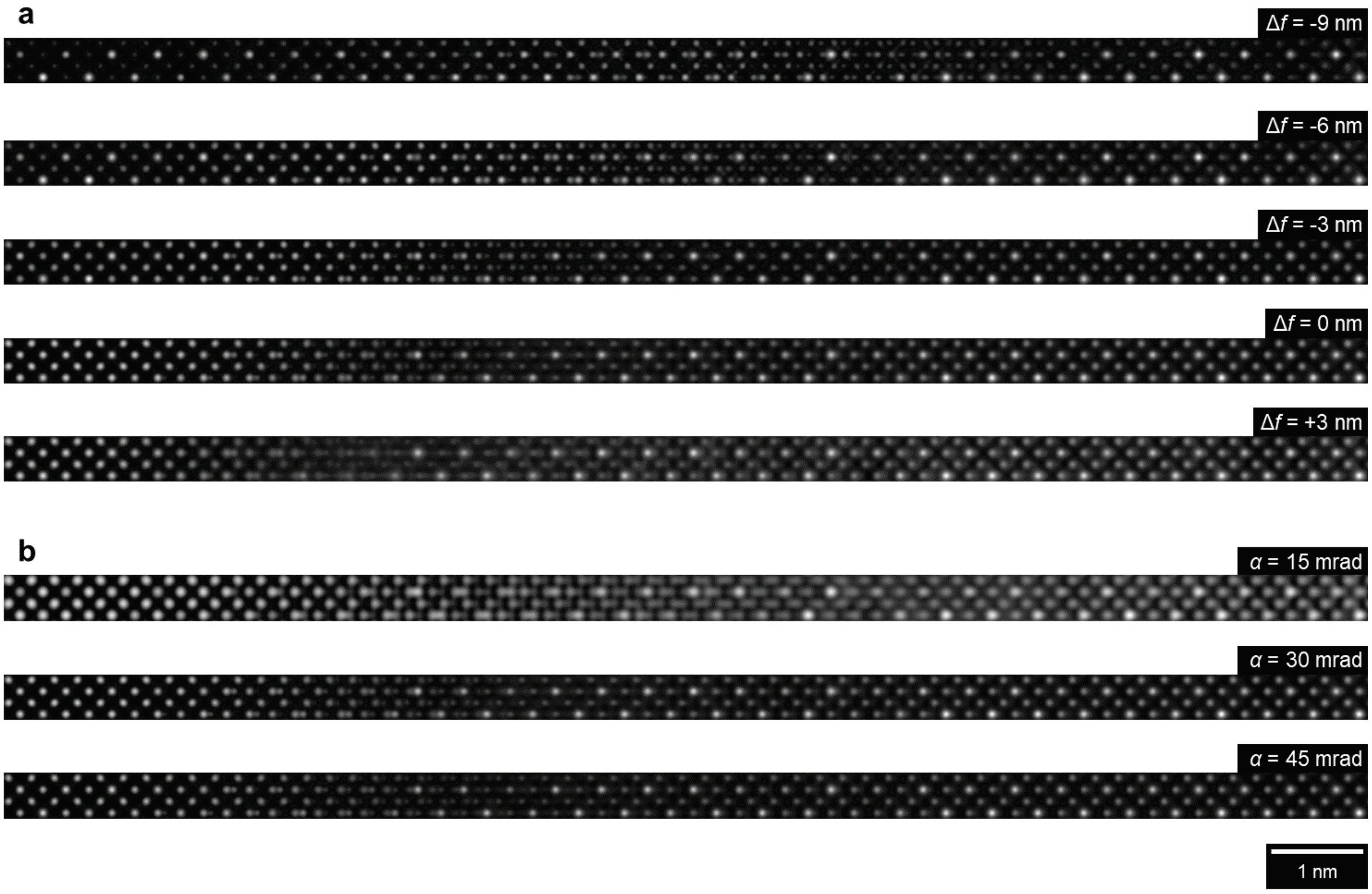

Supplementary Fig. 9: HAADF-STEM multislice simulations of an inclined SISF in $Co_3(Al,W)$ (a) under varying defocus $\Delta f$ at a constant beam semi-convergence angle of 30 mrad, and (b) under varying beam semi-convergence angle $\alpha$ at a constant defocus of 0 nm. Foil thickness ≈30 nm in all cases.

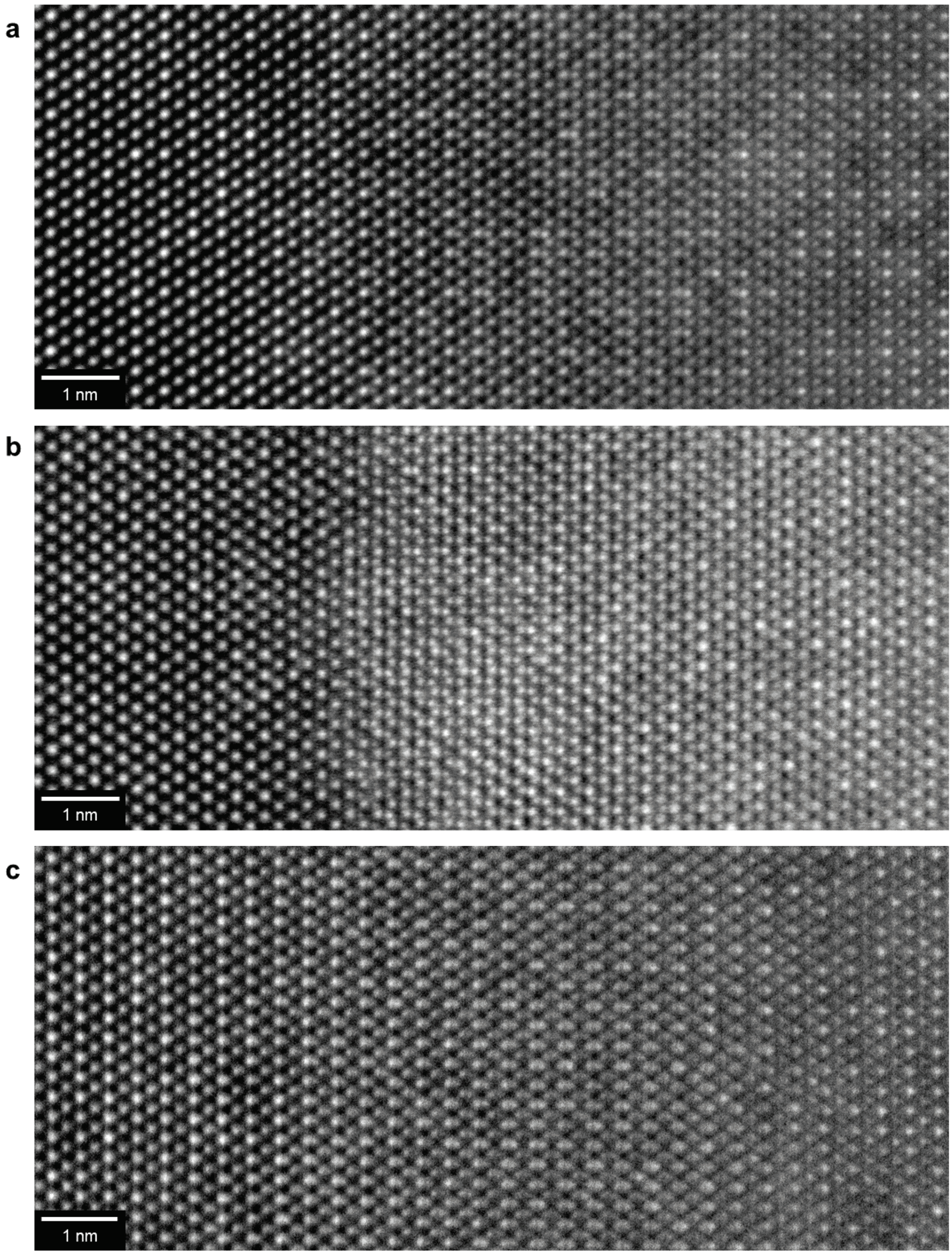

Supplementary Fig. 10: Influence of specimen preparation procedure on HRSTEM contrast of inclined SESFs. HRSTEM micrographs of inclined SFs in foils fabricated using different procedures: (a) electropolishing, (b) electropolishing with subsequent argon ion polishing; (c) focused ion-beam liftout. In all cases, individual atomic columns can be clearly resolved even in the visible overlap region between reference and shifted lattice. The foil thickness is between 60 and 70 nm in all cases.

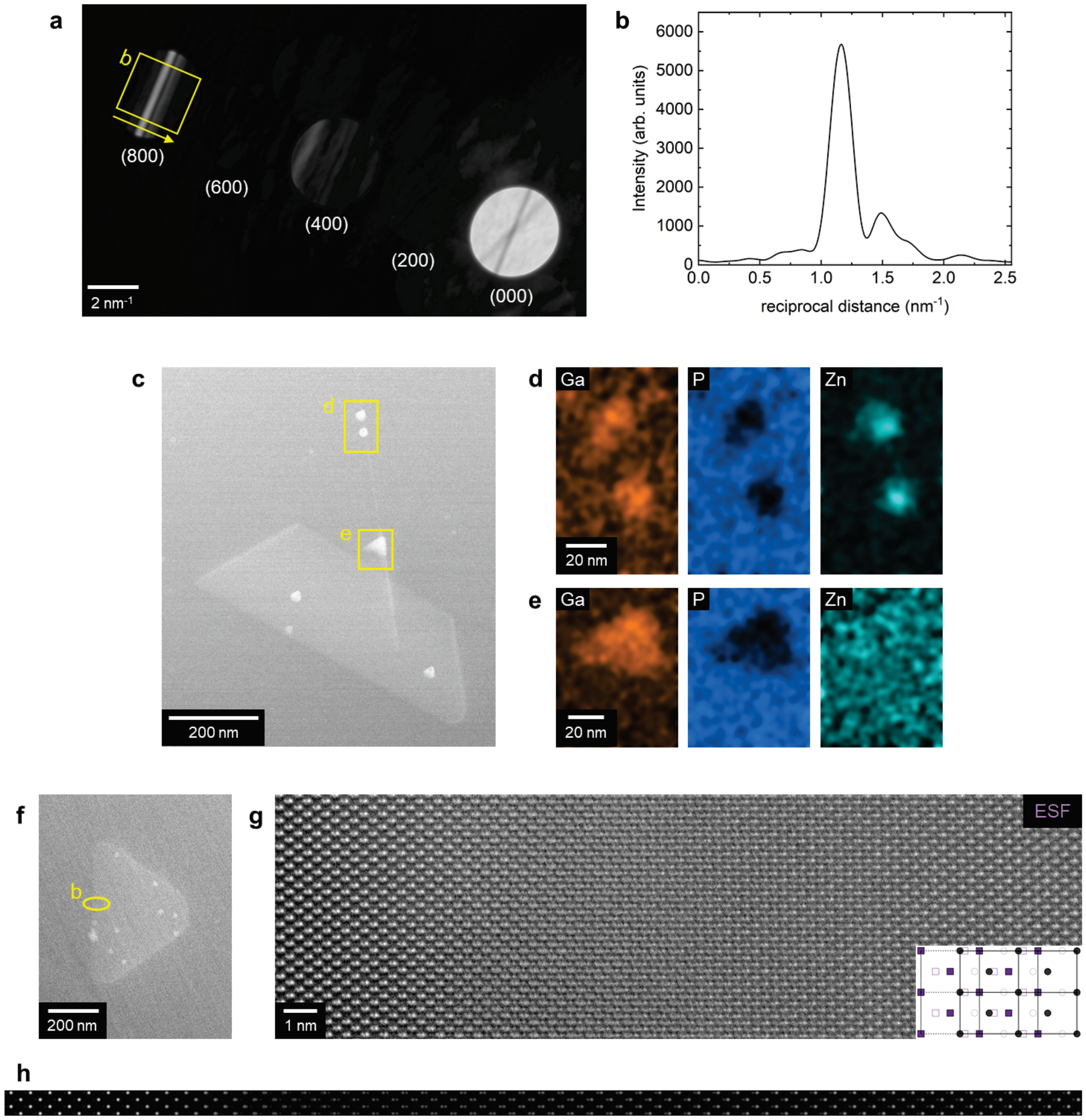

Supplementary Fig. 11: Additional results from the GaP specimen. (a) Energy-filtered CBED pattern with the beam converged on the SF shown in Fig. 8c in the main text, and the (800) reflection excited. (b) Intensity profile of the region marked in (a), clearly revealing splitting of the (800) line. The additional line of lower intensity appears in the direction of positive excitation error, indicating a phase shift of $+{}^{2\pi}/_{3}$ and thus an extrinsic fault. (c) ADF-STEM micrograph of the investigated specimen region. (d,e) EDXS elemental distribution maps (at.%) of the regions marked in (c), showing the elemental distribution in tetragonal voids partially filled with Ga and Zn. (f) Specimen region at a higher thickness of approx. 270 nm with an inclined SF. (g) HRSTEM micrograph of the region marked in (f), revealing the SF to be extrinsic. (h) HAADF-STEM multislice simulation of an inclined ESF in GaP at a thickness of ≈38 nm.

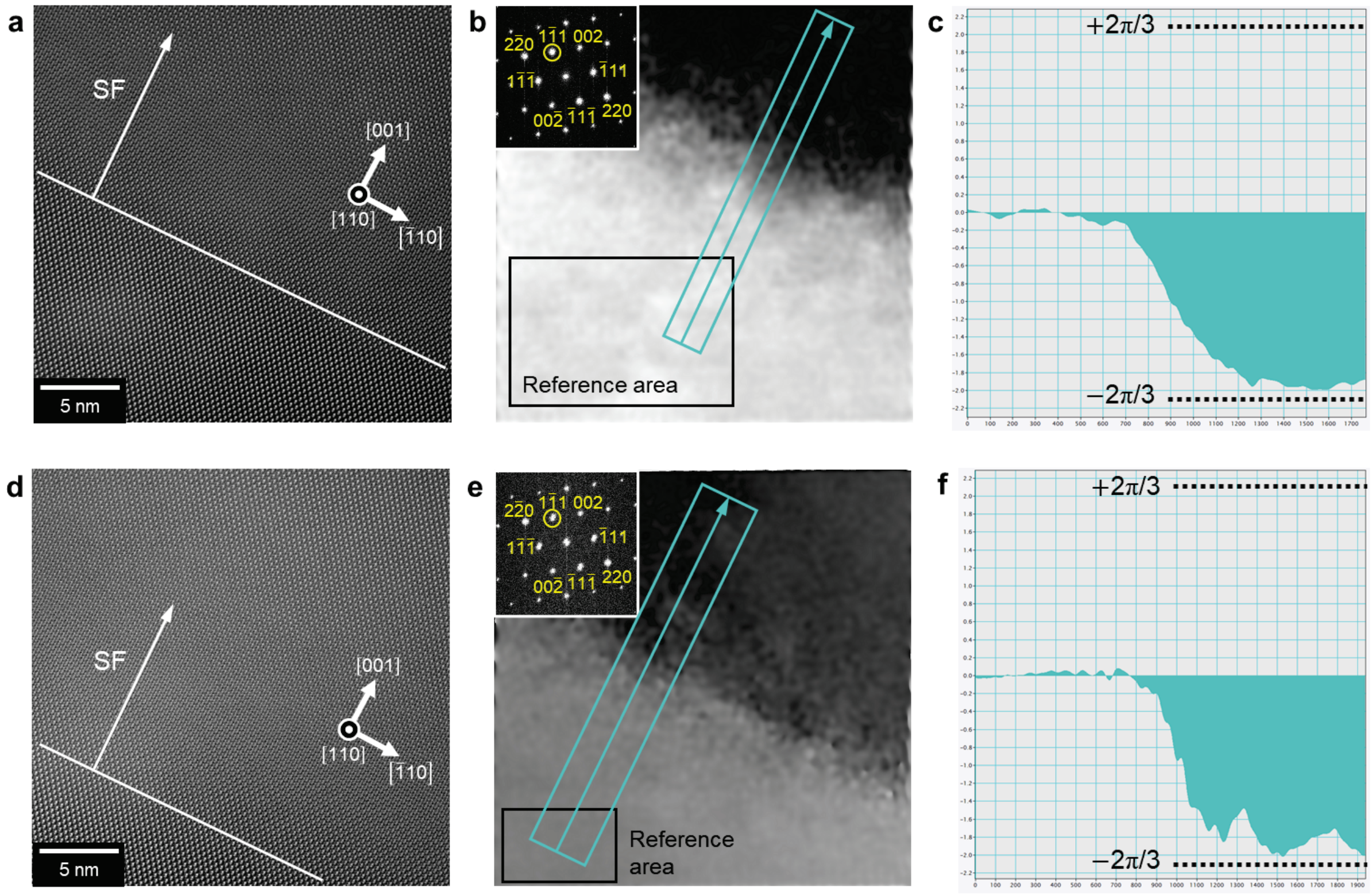

Supplementary Fig. 12: Application of GPA to determine SF type, exemplified by two extrinsic SFs in GaP. (a) HRSTEM micrograph of the inclined SF shown in Fig. 8c in the main text. The top edge of the SF is indicated by the white line. (b) Phase image created with the $(1\bar{1}1)$ reflection as indicated in the Fourier transform of (a) (see inset). The reference area, where the phase shift is defined to be zero, is chosen in the region of perfect lattice before the SF. Across the overlap region, a phase change is visible. (c) Line scan across the overlap region as highlighted in (b), indicating a phase shift towards $-{}^{2\pi}/_{3}$ and thus an extrinsic SF. (d) HRSTEM micrograph of the inclined SF shown in Supplementary Fig. 11g at a foil thickness of approx. 270 nm. (e,f) GPA analogous to the fault above; this SF is also revealed to be extrinsic.

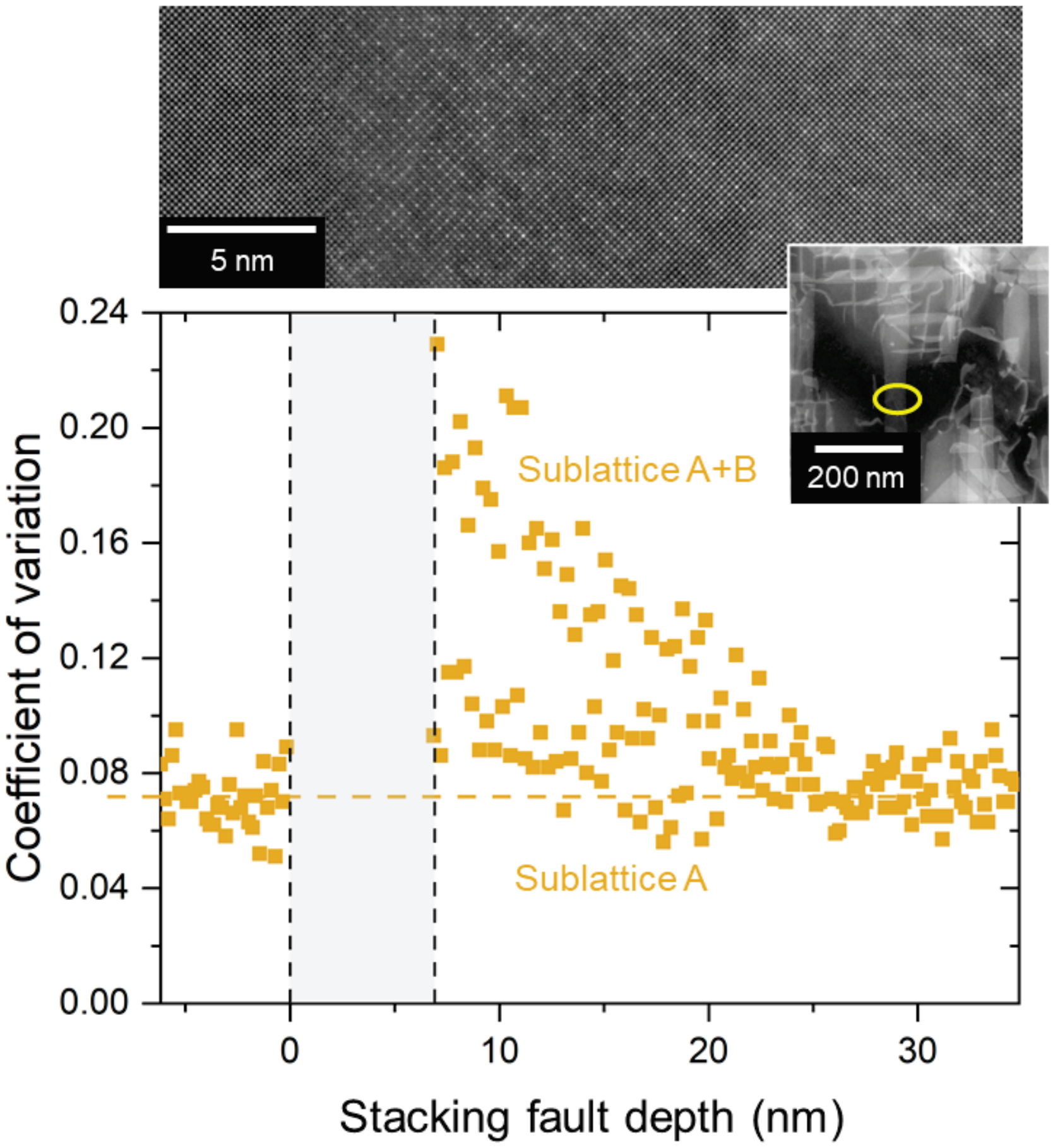

Supplementary Fig. 13: Analysis of enhanced Z-contrast above an inclined SF with a nano-γ′ precipitate. HRSTEM micrograph of an inclined SF in the fcc structure. The plot below shows the CV of atomic-column intensities within each (220) plane. The horizontal line represents the baseline SD determined in a region of perfect lattice to the left of the SF. The HRSTEM image was stitched together from two separate micrographs to obtain a field of view containing regions to the left and right of the inclined SF, from the region shown in the ADF-STEM micrograph (see inset). The nano-γ′ precipitate has a visible width of ≈15 nm.

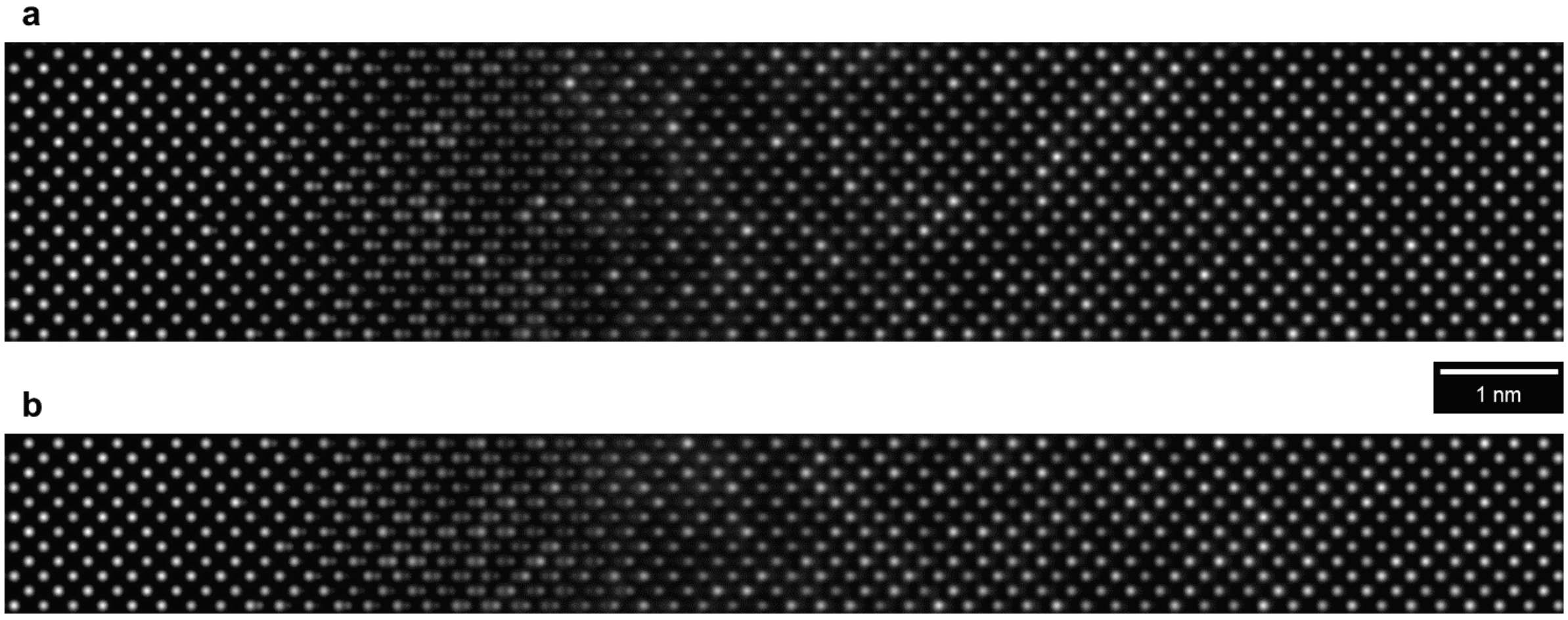

Supplementary Fig. 14: HAADF-STEM multislice simulations of inclined SFs in fcc solid solutions corresponding to the approximate compositions of the γ phase in (a) CoNi-base superalloy ERBOCo-4 ($Co_{50}Ni_{28}W_6Cr_9Ta_2Al_4Ti_1$ in at.%) and (b) Ni-base superalloy CMSX-4 ($Ni_{49}Co_{21}Cr_{18}W_4Re_3Al_4Mo_1$ in at.%). Foil thickness ≈20 nm in both cases.